\documentclass[12pt]{article}
\pdfoutput=1
\usepackage{amssymb}
\usepackage{amsmath}
\usepackage{latexsym}
\usepackage[usenames]{color}
\usepackage{fancybox}
\usepackage{simplewick}
\usepackage{comment}
\usepackage{cite}
\usepackage{framed}
\definecolor{shadecolor}{rgb}{0.9,0.9,0.95}
\usepackage{setspace}
\definecolor{darkgreen}{rgb}{0,0.5,0}
\definecolor{darkblue}{cmyk}{0.9,0.9,0,0}
\definecolor{darkred}{rgb}{0.6,0,0.3}

\usepackage{graphicx} 
\usepackage[setpagesize=false,pagebackref=false, linktocpage, bookmarksopen=true, colorlinks=true, linkcolor=darkblue,citecolor=darkblue,urlcolor=black]{hyperref}
\usepackage{hyperref}

\renewcommand{\thefootnote}{\arabic{footnote}}

\def\del{\partial}

\def\fn#1{\footnote{#1}}

\def\eqref#1{(\ref{#1})}
\def\comma{\,,}
\def\period{\,.}
\def\llangle{\langle \!\langle}
\def\rrangle{\rangle\!\rangle}
\def\beq{\begin{equation}}
\def\eeq{\end{equation}}

\begin{document}
\thispagestyle{empty}

\renewcommand{\thefootnote}{\fnsymbol{footnote}}
\setcounter{page}{1}
\setcounter{footnote}{0}
\setcounter{figure}{0}
\begin{flushright}
{\sf UT-Komaba/18-7}
\end{flushright}
\begin{center}
$$$$
{\Large\textbf{\mathversion{bold}
Correlation Functions on the Half-BPS Wilson Loop: Perturbation and Hexagonalization
}\par}

\vspace{1.3cm}

\textrm{Naoki Kiryu$^{\textcolor[rgb]{0.8,0,0.1}{\blacktriangleright}}$, Shota Komatsu$^{\textcolor[rgb]{0,0,0.8}{\blacktriangledown}}$}
\\ \vspace{1.2cm}
\footnotesize{\textit{
$^{\textcolor[rgb]{0.8,0,0.1}{\blacktriangleright}}$Institute of Physics, University of Tokyo, Komaba, Meguro-ku, Tokyo 153-8902 Japan\\
$^{\textcolor[rgb]{0,0,0.8}{\blacktriangledown}}$Perimeter Institute for Theoretical Physics, Waterloo, Ontario N2L 2Y5, Canada\\\&\\School of Natural Sciences, Institute for Advanced Study, Princeton, New Jersey 08540, USA\\
}  
\vspace{4mm}
}
\textrm{E-mail: {\tt kiryu AT hep1.c.u-tokyo.ac.jp, shota.komadze AT gmail.com}}

\par\vspace{1.5cm}

\textbf{Abstract}\vspace{2mm}
\end{center}
We compute correlation functions of protected primaries on the $1/2$-BPS Wilson loop in $\mathcal{N}=4$ super Yang-Mills theory at weak coupling. We first perform direct perturbative computation at one loop in the planar limit and present explicit formulae for general two-, three- and four-point functions. The results for two- and three-point functions as well as four-point functions in special kinematics are in perfect agreement with the localization computation performed in \href{http://arxiv.org/abs/1802.05201}{\tt arXiv:1802.05201}. We then analyze the results in view of the integrability-based approach called ``hexagonalization'', which was introduced previously to study the correlation functions in the absence of the Wilson loop. In this approach, one decomposes the correlator into fundamental building blocks called ``hexagons'', and glues them back together by summing over the intermediate states. Through the comparison, we conjecture that the correlation functions on the Wilson loop can be computed by contracting hexagons with boundary states, where each boundary state represents a segment of the Wilson loop. As a byproduct, we make predictions for the large-charge asymptotics of the structure constants on the Wilson loop. Along the way, we refine the conjecture for the integrability-based approach to the general non-BPS structure constants on the Wilson loop, proposed originally in \href{http://arxiv.org/abs/1706.02989}{\tt arXiv:1706.02989}.  
\noindent

\setcounter{page}{1}
\renewcommand{\thefootnote}{\arabic{footnote}}
\setcounter{footnote}{0}
\setcounter{tocdepth}{2}
\newpage
\tableofcontents

\parskip 5pt plus 1pt   \jot = 1.5ex

\newpage
\section{Introduction and Summary\label{sec:intro}}
The idea of reformulating gauge theories as string theory dates back at least to an early attempt by Polyakov \cite{Polyakov:1980ca} in which he tried to re-interpret the gauge theories by treating the color-electric fluxes, or equivalently the Wilson loops, as fundamental degrees of freedom. A closely-related idea was explored at around the same time by Migdal and Makeenko who wrote down the Schwinger-Dyson equation for the Wilson loop, known as the loop equation \cite{Makeenko:1979pb,Makeenko:1980vm}.

The scene has changed dramatically after the discovery of the AdS/CFT correspondence \cite{Maldacena:1997re}. It provided the first concrete example of the equivalence between gauge and string theories and gave a clear physical meaning to 't Hooft's planar expansion \cite{tHooft:1973alw}: Namely, the two-dimensional surfaces that appear in the 't Hooft expansion are interpreted as the worldsheets of {\it physical} strings living in anti-de-Sitter space. Besides being a conceptual breakthrough, the correspondence led to new computational tools since it enabled us to analyze the higher-dimensional gauge theories using the techniques of the two-dimensional field theories. The most notable success in this regard is the application of integrability to $\mathcal{N}=4$ supersymmetric Yang-Mills theory (SYM) in four dimensions \cite{Minahan:2002ve,Beisert:2010jr}. As a result of relentless efforts in the past fifteen years, the integrability of the worldsheet was applied successfully to the computation of the spectrum \cite{Gromov:2009tv,Bombardelli:2009ns, Arutyunov:2009ur,Gromov:2013pga}, the scattering amplitudes \cite{Basso:2013vsa}, the structure constants \cite{Basso:2015zoa}, and the four- \cite{Caetano:2011eb,Eden:2016xvg,Caetano:2012ac,Fleury:2016ykk} and higher-point correlators \cite{Fleury:2016ykk,Fleury:2017eph} in the planar limit and their first nonplanar corrections \cite{Ben-Israel:2018ckc,Bargheer:2017nne,Eden:2017ozn,Bargheer:2018jvq}.

In this paper, we analyze quantities that might help us make connections between these old and new stories. More specifically, we study the correlation functions of operator insertions on the Wilson loop at weak coupling. On the one hand, they can be viewed as simple generalization of standard correlation functions of local operators and are likely to be amenable to the integrability machinery as we discuss in this paper. On the other hand, the operator insertions on the Wilson loop describe the local deformation of the loop and are therefore closely tied to the loop equation. Although what we will achieve in this paper is perhaps less than the first step, we hope that our study will pave the way towards the future developments and help to clarify the relation between the old ideas, such as the loop equation, and the modern viewpoint on the gauge-string duality.
\subsection{Correlators on the $1/2$-BPS Wilson loop}
Having explained the underlying motivation, let us now describe more in detail the contents of this paper. Our target is the simplest but perhaps the most basic of all possible correlators on the Wilson loops; the correlators of protected primaries on the $1/2$-BPS Wilson loop in $\mathcal{N}=4$ SYM. The $1/2$-BPS Wilson loop is a supersymmetric generalization of the ordinary Wilson loop  which preserves the largest amount of supersymmetries. It couples to a scalar as well as to a gauge field and takes the form,
\beq
\mathcal{W}=\frac{1}{N}{\rm Tr}\left[{\rm P} \exp \oint_{\mathcal{C}} \left(i A_\mu \dot{x}^{\mu} +\Phi_6|\dot{x}|\right)ds\right]\comma
\eeq
where the contour $\mathcal{C}$ can be a circle or a straight line. It has several distinctive features as compared to other Wilson loops. First, thanks to the coupling to the scalar, it is manifestly UV-finite. Second, because it preserves a large amount of supersymmetries, its expectation value can be computed by the supersymmetric localization \cite{Erickson:2000af,Drukker:2000rr,Pestun:2007rz}. Third, since its contour is circular or a straight line, it preserves the SL$(2,R)$ subgroup of the full conformal group and therefore can be regarded as a one-dimensional conformal defect \cite{Billo:2016cpy,Liendo:2016ymz,Liendo:2018ukf}. This last point provides an incentive to study the correlators of operator insertions,
\beq
\langle \mathcal{W}[\mathcal{O}_1 (x_1)\cdots \mathcal{O}_n (x_n)]\rangle \equiv \left< \frac{1}{N}{\rm Tr}\left[{\rm P} \left( \mathcal{O}_1 (x_1)\cdots \mathcal{O}_n (x_n)e^{\oint_{\mathcal{C}} \left(i A_\mu \dot{x}^{\mu} +\Phi_6|\dot{x}|\right)ds}\right)\right]\right>\comma
\eeq
 since they are the most natural observables of this defect CFT. Such defect correlation functions\fn{For other related works which analyzed the $1/2$-BPS Wilson loop in $\mathcal{N}=4$ SYM from the defect CFT perspective, see \cite{Kim:2017sju,Cavaglia:2018lxi,Giombi:2018qox,Giombi:2018hsx}. For the analysis on the non-supersymmetric Wilson loops (from the defect CFT and from integrability), see \cite{Beccaria:2017rbe,Beccaria:2018ocq,Correa:2018fgz}.} were studied in detail at strong coupling by analyzing the fluctuations around a classical string configuration in AdS$_2$ \cite{Giombi:2017cqn} while at weak coupling they were studied using the deformation of the Wilson loop in \cite{Cooke:2017qgm,Cooke:2018obg}. The spectrum, or equivalently the two-point functions, of operator insertions were computed from integrability in \cite{Drukker:2006xg,Drukker:2012de,Correa:2012hh}\fn{It was later reformulated in an elegant way by the Quantum Spectral Curve \cite{Gromov:2015dfa}.}. The focus of this paper is on the three- and four-point functions, and in particular on the special operator insertions
 \beq
 \mathcal{O}_m(x_m)= (Y_m \cdot \Phi)^{L_m}(x_m)\comma
 \eeq 
 where $Y_m$ is a five-dimensional null vector and $Y \cdot \Phi$ signifies the inner product $\sum_{\mu=1}^{5}Y^{\mu} \Phi_{\mu}$. Owing to the supersymmetries, the dimensions of these operators are tree-level exact while their three- and four-point functions do depend on the coupling constant.
 
Precisely speaking, to make contact with the defect CFT data, we should consider the {\it normalized} correlator which is obtained by dividing the correlator by the expectation value of the Wilson loop as
 \beq
 \llangle  \mathcal{O}_1(x_1)\cdots \mathcal{O}_n(x_n) \rrangle\equiv \frac{\langle \mathcal{W}[\mathcal{O}_1 (x_1)\cdots \mathcal{O}_n (x_n)]\rangle}{\langle \mathcal{W}\rangle}\period 
 \eeq
For the straight-line Wilson loop, this manipulation is trivial since the expectation value of the Wilson loop is unity while for the circular Wilson loop it involves the division by the planar expectation value \cite{Erickson:2000af,Drukker:2000rr},
\beq
\langle \mathcal{W}\rangle_{\rm circle} =\frac{2}{\sqrt{\lambda}}I_1 (\sqrt{\lambda})\comma
\eeq
with $\lambda$ being the `t Hooft coupling constant $\lambda\equiv g_{\rm YM}^2 N$.
 The spacetime dependence of these correlators is constrained by the SL$(2,R)$ invariance. For instance, the two- and the three-point functions are given by
 \beq
 \begin{aligned}
 \llangle\mathcal{O}_1(x_1) \mathcal{O}_2(x_2) \rrangle&=n_{L_1}\delta_{L_1,L_2}  \times (d_{12})^{L_1}\comma\\
 \frac{\llangle \mathcal{O}_1(x_1) \mathcal{O}_2(x_2) \mathcal{O}_3(x_3)\rrangle}{\sqrt{n_{L_1}n_{L_2}n_{L_3}}}&= \frac{C_{L_1,L_2,L_3}}{\sqrt{N}}\times (d_{12})^{\frac{L_{12|3}}{2}}(d_{23})^{\frac{L_{23|1}}{2}}(d_{31})^{\frac{L_{31|2}}{2}}\comma
 \end{aligned}
 \eeq
  where $N$ is the rank of the gauge group, $L_{ij|k}\equiv L_i+L_j-L_k$ and $d_{ij}$ is the free-field Wick contraction which takes the following form for the straight-line Wilson loop:
  \beq\label{eq:dijstraight}
  \left.d_{ij}\right|_{\text{straight line}}= \frac{Y_i \cdot Y_j}{x_{ij}^2}\comma\qquad  \qquad x_{ij}^2\equiv |x_i-x_j|^2\period
  \eeq
  The quantity $c_{L_1,L_2,L_3}$ is the structure constant of the defect CFT while $n_{L}$ is the normalization of the two-point function, which can be set to unity by rescaling the operators\fn{The protected primaries can be obtained by taking the OPE of single-scalar insertions $\Phi_i$ $(i\neq 6)$, which are in the same supermultiplet as the displacement operator and have canonical normalization. Therefore it is possible to attribute certain physical meaning to the normalization $n_{L}$ (see for instance \cite{Giombi:2018qox}). However, we will not advocate this point of view in this paper.}. For the circular Wilson loop, one just needs to replace $d_{ij}$ with
\beq\label{eq:dijcircle}
  \left.d_{ij}\right|_{\rm circle}=\frac{Y_i\cdot Y_j}{(2\sin \frac{\tau_i-\tau_j}{2})^2}\comma
\eeq
  where $\tau_i$ parameterizes the position of the insertion on the circle and ranges from $0$ to $2\pi$. Note that the two expressions \eqref{eq:dijstraight} and \eqref{eq:dijcircle} are related by the conformal transformation.
  
 On the other hand, the four-point function 
  \beq
  \frac{\llangle\mathcal{O}_1 (x_1)\mathcal{O}_2 (x_2) \mathcal{O}_3 (x_3)\mathcal{O}_4 (x_4) \rrangle}{\sqrt{n_{L_1}n_{L_2}n_{L_3}n_{L_4}}}\equiv \frac{G_{L_1,L_2,L_3,L_4}}{N}\comma
  \eeq
  is a nontrivial function of the cross ratios. To see this explicitly, we strip off the space-time (and the R-symmetry) dependence from $G_{L_1,L_2,L_3,L_4}$ as
  \beq
  G_{L_1,L_2,L_3,L_4} = d_{12}^{\frac{L_1+L_2}{2}}d_{34}^{\frac{L_3+L_4}{2}}\left(\frac{d_{24}}{d_{14}}\right)^{\frac{L_2-L_1}{2}}\left(\frac{d_{13}}{d_{14}}\right)^{\frac{L_3-L_4}{2}}g_{L_1,L_2,L_3,L_4}(\chi,\alpha,\bar{\alpha}) \period
  \eeq
  Then, the remaining quantity $g_{L_1,L_2,L_3,L_4}$ depends only on the cross ratios, defined by\fn{Here we presented the definition for the straight-line Wilson loop. For the circular Wilson loop, one needs to replace $x_{ij}^2$ with $(2\sin \frac{\tau_i-\tau_j}{2})^2$.}
  \beq\label{eq:def1dCR}
  \begin{aligned}
  &\frac{x_{12}^2x_{34}^2}{x_{13}^2x_{24}^2}=\chi^2\comma\qquad &&\frac{x_{14}^2x_{23}^2}{x_{13}^2x_{24}^2}=(1-\chi)^2\comma\\
  &\frac{(Y_1\cdot Y_2)(Y_3\cdot Y_4)}{(Y_1\cdot Y_3)(Y_2\cdot Y_4)}=\alpha \bar{\alpha}\comma\qquad &&\frac{(Y_1\cdot Y_4)(Y_2\cdot Y_3)}{(Y_1\cdot Y_3)(Y_2\cdot Y_4)}=(1-\alpha)(1- \bar{\alpha})\period
  \end{aligned}
  \eeq
 
 Let us now make one important remark: In one-dimensional (defect) CFTs, one should be careful about the ordering of the operators since the correlators with different orderings are not related by a simple analytic continuation even in the Euclidean kinematics. This is in marked contrast to the higher-dimensional CFTs in which one can continuously move one operator around another to reach a different configuration. In terms of the conformal cross ratio, the different orderings correspond to different ranges of $\chi$ as\fn{These are all possible orderings in the presence of the parity invariance.}
 \beq
 \begin{aligned}
&\{ 2134\}:\,\, \chi \in [-\infty,0]\comma \qquad \{1234\}: \,\,\chi \in [0,1]\comma \qquad \{1324\}:\,\, \chi \in [1,\infty]\comma 
 \end{aligned}
 \eeq
 where $\{ ijkl\}$ signifies the correlator with the operator ordering $\mathcal{O}_i \mathcal{O}_j \mathcal{O}_k \mathcal{O}_l$. Thus the above statement translates to the fact that the correlators with different values of $\chi$ are not simply related by the analytic continuation. In the rest of this paper, to avoid any possible confusion arising from this point, we always consider the correlators in the ordering $\{1234\}$. In other words, we always assume that the cross ratio $\chi$ takes the value between $0$ and $1$. 
\subsection{Outline of the paper and brief summary of the results}
In section \ref{sec:perturbation}, we compute two-, three- and four-point functions at one loop from perturbation theory. To simplify the analysis, we consider the correlators on the straight line. As explained above, such correlators have direct relation to the defect CFT data, see \eqref{eq:3ptfinal} and \eqref{eq:oneloopfinal2} for final results. The results for the three-point functions are in perfect agreement with the exact results obtained previously by localization \cite{Giombi:2018qox} while the results for the four-point functions are new except in special kinematics in which they correctly reduce to the results from localization in \cite{Giombi:2018qox}.

\begin{figure}[t]
\centering
\includegraphics[clip,height=4cm]{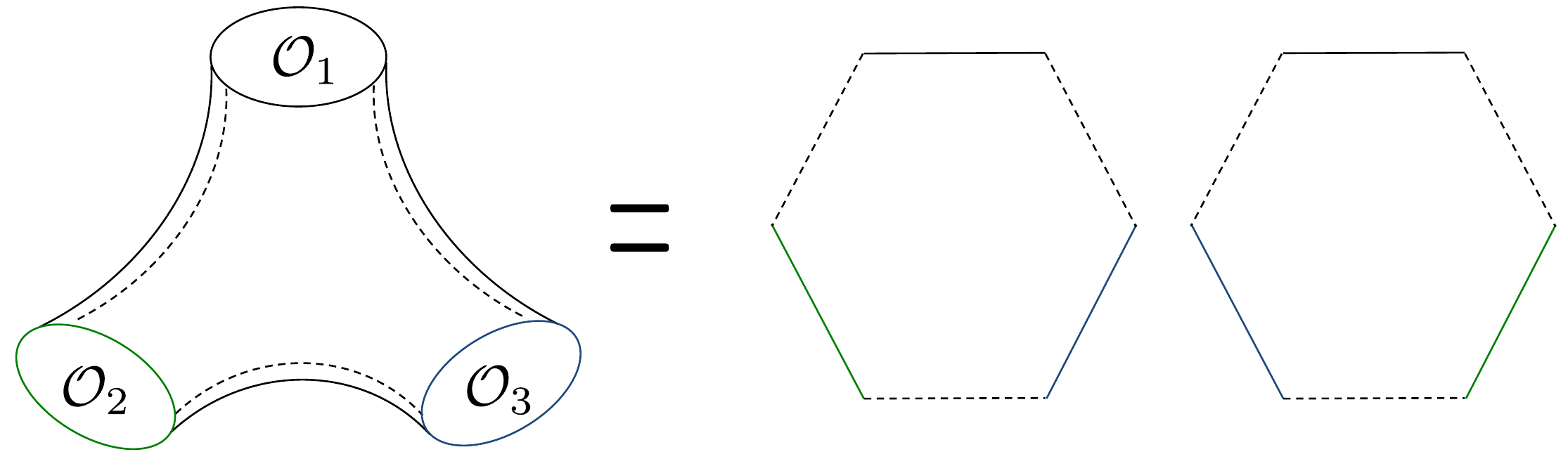}
\caption{Hexagonalization of a three-point function of single-trace operators. The string worldsheet dual to the three-point function has a pair-of-pants topology. The idea of the hexagonalization is to cut the pair of pants into two hexagonal patches and stitch them together by summing over intermediate states living on the dashed edges.}
\label{fig:fig0}
\end{figure}
\begin{figure}[t]
\centering
\includegraphics[clip,height=4cm]{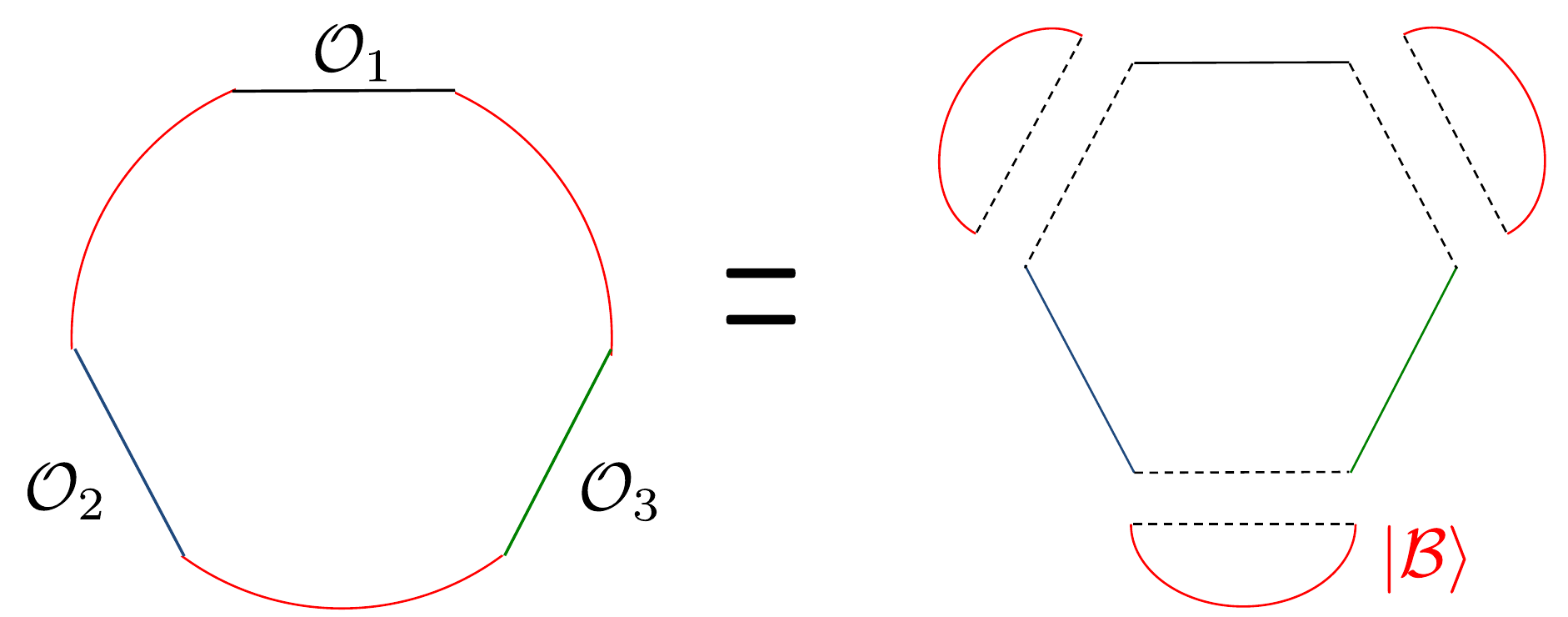}
\caption{Hexagonalization of a three-point function on the Wilson loop. The three-point function on the Wilson loop can be decomposed into a single hexagon and three boundary states. To glue the boundary states to the hexagon, we contract the boundary states $|\mathcal{B}\rangle$, which are denoted by red curves and correspond to segments of the Wilson loop, with the hexagon.}
\label{fig:fig1}
\end{figure}

In section \ref{sec:hexagonalization}, we examine these results from the point of view of integrability. In particular, we compare the results with the general expectation in the ``hexagonalization'' approach \cite{Fleury:2016ykk}: The basic idea of the hexagonalization is to view the correlator as a worldsheet of string and decompose it into smaller hexagonal patches called {\it hexagons} \cite{Basso:2015zoa}. After the decomposition, one computes the contribution from each hexagon using integrability and glues them back together by summing over the intermediate states that appear on the glued edges (see figure \ref{fig:fig0}). Through the comparison, we conjecture that the three-point function on the Wilson loop can be decomposed into one hexagon and three boundary states  (see figure \ref{fig:fig1}), where the relevant boundary state was determined in the study of the spectrum on the Wilson loop \cite{Drukker:2012de,Correa:2012hh}. The idea of computing the three-point functions on the Wilson loop from hexagons was proposed originally in \cite{Kim:2017phs}. However, our analysis suggests the necessity of the refinement of the proposal \cite{Kim:2017phs}, which we will describe shortly. 

We also analyze the four-point function in a similar fashion; namely we propose that it can be computed by gluing two hexagons and four boundary states and multiplying a nontrivial cross-ratio-dependent weight factor, which was first determined in the study of correlators of single-trace operators in \cite{Fleury:2016ykk}.

One interesting (and perhaps peculiar) outcome of our analysis is that the integrability computation does not seem to correspond to the normalized correlator $\llangle \ast\rrangle$. It rather corresponds to the ratio of the correlators on the circular Wilson loop,
\beq\label{eq:proposalhex}
{\tt Hexagonalization}=\frac{\langle \mathcal{W} [\mathcal{O}_1\cdots \mathcal{O}_m]\rangle_{\rm circle}}{\prod_{k=1}^{m}\sqrt{\langle \mathcal{W}[\mathcal{O}_k\mathcal{O}_k]\rangle_{\rm circle}^{\prime}}}\comma
\eeq
where $\langle \mathcal{W}[\mathcal{O}_k\mathcal{O}_k]\rangle_{\rm circle}^{\prime}$ denotes the space-time independent part of the two-point function on the circular loop; namely $\langle \mathcal{W}[\mathcal{O}_k\mathcal{O}_k]\rangle_{\rm circle}^{\prime} =n_{L_k}\times \langle\mathcal{W} \rangle_{\rm circle}$. In terms of normalized correlators, the conjecture \eqref{eq:proposalhex} can be rewritten as
\beq
{\tt Hexagonalization}=\left(\langle \mathcal{W} \rangle_{\rm circle}\right)^{\frac{2-m}{2}} \frac{\llangle\mathcal{O}_1\cdots \mathcal{O}_m \rrangle}{\sqrt{\prod_{k=1}^{m}n_k}}\period
\eeq
Although we do not have a strong argument as to why the hexagonalization computes the correlators on the circle rather than the normalized correlators, we show that this conjecture is consistent with the one-loop three- and four-point functions we compute. Furthermore, it is supported by the comparison with the localization results at higher loops. In addition, we show that the conjecture leads to the following large-charge asymptotics for the structure constant on the Wilson loop,
\beq
C_{L_1,L_2,L_3}\quad \overset{L_k\to\infty}{\sim}\quad (\langle \mathcal{W}\rangle_{\rm circle})^{\frac{1}{2}}=\left(\frac{2I_1 (\sqrt{\lambda})}{\sqrt{\lambda}}\right)^{\frac{1}{2}}\period
\eeq

Based on this observation, in section \ref{sec:nonBPS3pt} we present a refined version of the conjecture on the hexagon formalism for the non-BPS structure constants on the $1/2$-BPS Wilson loop, which was given originally in \cite{Kim:2017phs}. Finally in section \ref{sec:conclusion}, we conclude and comment on future directions. Several appendices are included to explain technical details.
\section{Perturbation I: Two-and Three-Point Functions\label{sec:perturbation}}
In this section, we describe our one-loop computation of three- and four-point functions.
\subsection{Building blocks}
Before analyzing the correlators, it is useful to collect the results of various diagrams that appear in the computation.
\paragraph{Conformal Integrals}
Before analyzing individual diagrams, let us first list the relevant conformal integrals. The most important one is the one-loop conformal integral which is defined by
\beq
\Phi(z,\bar{z})=\frac{x_{13}^2x_{24}^2}{\pi^2}\int \frac{d^4 x_5}{x_{15}^2x_{25}^2x_{35}^2x_{45}^2}\comma
\eeq
where $z$ and $\bar{z}$ are the conformal cross ratios given by
\beq
z\bar{z}=\frac{x_{12}^2x_{34}^2}{x_{13}^2x_{24}^2} \comma\qquad (1-z)(1-\bar{z})=\frac{x_{14}^2x_{23}^2}{x_{13}^2x_{24}^2}\period
\eeq
The integral can be evaluated explicitly and we get
\beq
\Phi(z,\bar{z})=\frac{2{\rm Li}_2 (z)-2{\rm Li}_2 (\bar{z})+\log z\bar{z}\log \frac{1-z}{1-\bar{z}}}{z-\bar{z}}\period
\eeq

Another important integral is the three-point conformal integral defined by
\beq
\mathcal{Y}_{123}=\int \frac{d^4 x_5}{x_{15}^2x_{25}^2x_{35}^2}\period
\eeq
Using the fact that the three-point integral can be obtained as a limit of the four-point integral, we can compute this integral as
\beq
\mathcal{Y}_{123}=\lim_{x_4\to\infty}x_4^2 \int \frac{d^4 x_5}{x_{15}^2x_{25}^2x_{35}^2x_{45}^2}=\frac{\pi^2\Phi (z^{\prime},\bar{z}^{\prime})}{x_{13}^2}\comma
\eeq
with
\beq
z^{\prime}\bar{z}^{\prime}=\frac{x_{12}^2}{x_{13}^2}\comma\qquad (1-z^{\prime})(1-\bar{z}^{\prime})=\frac{x_{23}^2}{x_{13}^2}\period
\eeq

\begin{figure}[t]
\centering
\includegraphics[clip,height=4.5cm]{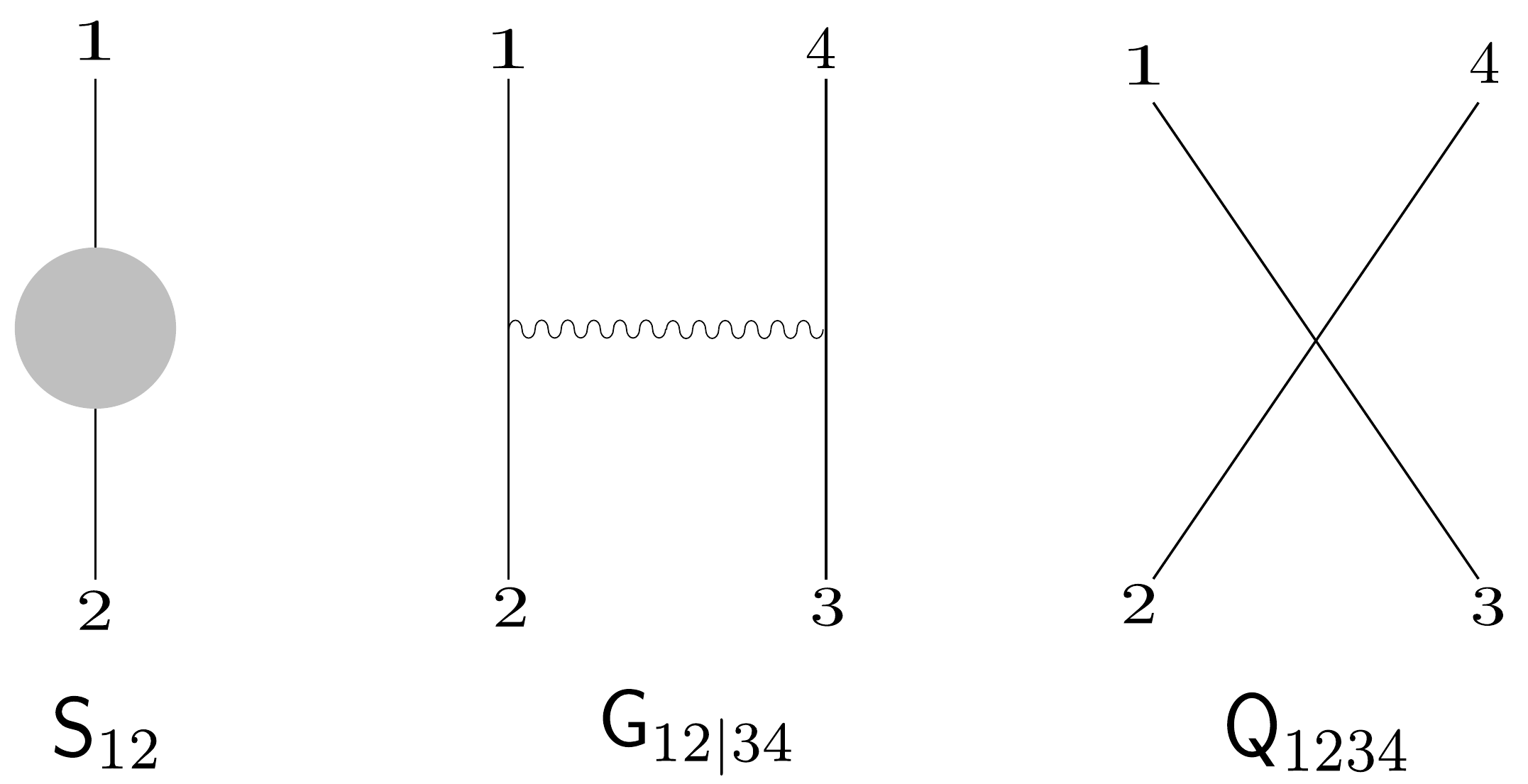}
\caption{Basic one-loop diagrams: The self-energy diagram ${\sf S}_{12}$, the gluon-exchange diagram ${\sf G}_{12|34}$ and the scalar quartic diagram ${\sf Q}_{1234}$.}
\label{fig:building}
\end{figure}

In the computation of correlators on the Wilson loop, all the points lie on a single line. In such a case, these conformal integrals simplify to
\beq\label{eq:explicitconformal}
\begin{aligned}
\Phi(\chi,\chi)&=-\left[\frac{\log \chi^2}{1-\chi}+\frac{\log (1-\chi)^2}{\chi}\right]\comma\\
\mathcal{Y}_{123}&=-2\pi^2\left(\frac{\log |x_{12}|}{x_{13} x_{23}}+\frac{\log |x_{13}|}{x_{12}x_{32}}+\frac{\log |x_{23}|}{x_{21}x_{31}}\right)\comma
\end{aligned}
\eeq
where $\chi$ is the one-dimensional conformal cross ratio defined by \eqref{eq:def1dCR}. In what follows, we often denote $\Phi(\chi,\chi)$ simply by $\Phi (\chi)$. The integral $\Phi (\chi)$ satisfies the following identities (see for instance \cite{Fleury:2016ykk})
\beq
\Phi (1-\chi)=\Phi (\chi)\comma\qquad \Phi \left(\frac{1}{\chi}\right)=\chi^2 \Phi (\chi)\comma\qquad \Phi \left(\frac{\chi}{\chi-1}\right)=(1-\chi)^2\Phi (\chi)\period
\eeq
\paragraph{Bulk Diagrams}
Let us first consider the diagrams that do not involve the contraction with the Wilson loop. The three basic building blocks for such diagrams are the self-energy diagram, the gluon exchange diagram, and the scalar quartic diagram (see figure \ref{fig:building}). These are the diagrams which also show up in the computation of correlators of single-trace operators and have been computed in the literature (see for instance \cite{Drukker:2008pi}). In what follows, the computations are done using the point-splitting regularization. In addition, we use the standard convention in the integrability literature and denote the coupling constant by $g$, which is related to the 't Hooft coupling constant $\lambda$ as follows:
\beq
g=\frac{\sqrt{\lambda}}{4\pi}\period
\eeq

The result for the self-energy diagram ${\sf S}_{ij}$ is
\beq\label{eq:selfenergy}
{\sf S}_{12}=-4g^2\left(\log \frac{|x_{12}|}{\epsilon}+1\right)\bar{d}_{12}
\eeq
where $\bar{d}_{ij}\equiv 2g^2 d_{ij}$.
Similarly, the results for the gluon exchange diagram ${\sf G}_{ij|kl}$ and the scalar quartic diagram ${\sf Q}_{ijkl}$ are
\beq\label{eq:defGQ}
\begin{aligned}
&{\sf G}_{12|34}=\frac{g^2}{2}\bar{d}_{12}\bar{d}_{34}\Phi (\chi)\left(\chi^2-2\chi\right)+{\sf C}_{[12][34]}\comma\\
&{\sf Q}_{1234}=\frac{g^2}{2}\Phi (\chi) \left[2\bar{d}_{13}\bar{d}_{24}-(1-\chi)^2\bar{d}_{23}\bar{d}_{14}-\chi^2\bar{d}_{12}\bar{d}_{34}\right]\comma
\end{aligned}
\eeq
with\fn{The function ${\sf C}_{[12][34]}$ is antisymmetric with respect to the exchange of $1$ and $2$ (or $3$ and $4$). This is the reason for writing the subscripts as $[12][34]$.}
\beq
\begin{aligned}
{\sf C}_{[12][34]}\equiv &\frac{g^2}{2}\bar{d}_{12}\bar{d}_{34}\times\\
&\frac{\left(x_{13}^2-x_{23}^2\right)\mathcal{Y}_{123}-\left(x_{14}^2-x_{24}^2\right)\mathcal{Y}_{124}+\left(x_{13}^2-x_{14}^2\right)\mathcal{Y}_{134}-\left(x_{23}^2-x_{24}^2\right)\mathcal{Y}_{234}}{\pi^2}\period
\end{aligned}
\eeq
Here and below the subscripts specify the pattern of the contractions (see figure \ref{fig:building} for the explicit diagrammatic definition of the corner contribution).
\begin{figure}[t]
\centering
\includegraphics[clip,height=2.7cm]{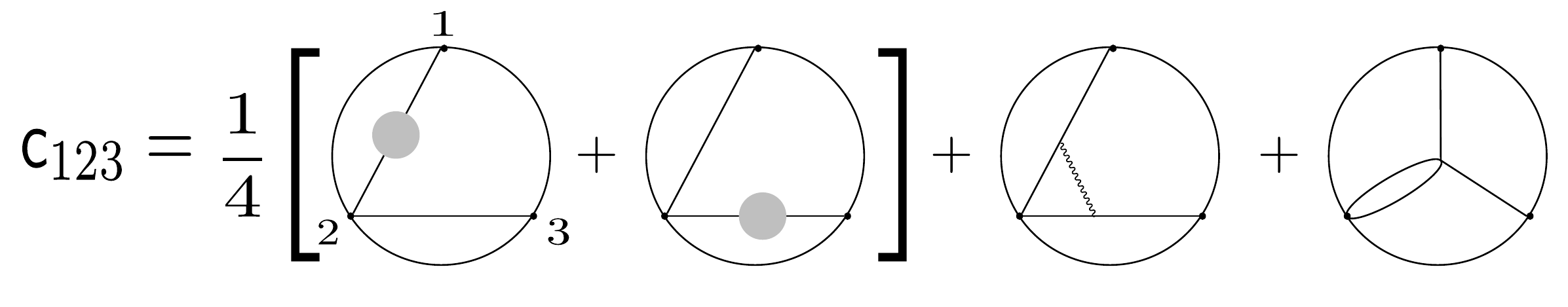}
\caption{Definition of the corner contribution ${\sf c}_{123}$. Although the figures are for the circular Wilson loop, the acutal computation was performed for the straight-line Wilson loop.}
\label{fig:corner}
\end{figure}

These are the basic sets of the ``bulk'' diagrams and other diagrams can be obtained by taking the limit of these diagrams. Particularly important among them is the corner contribution ${\sf c}_{ijk}$ \cite{Drukker:2008pi}, which is a weighted sum of diagrams that are associated with a ``corner'' of the planar Wick contraction (see figure \ref{fig:corner}): 
\beq
{\sf c}_{123}\bar{d}_{12}\bar{d}_{23}\equiv\frac{1}{4}\left({\sf S}_{12}\bar{d}_{23}+\bar{d}_{12}{\sf S}_{23}\right)+{\sf G}_{12|23}+{\sf Q}_{1223} \period
\eeq
Using \eqref{eq:defGQ}, one can verify that the result can be simply written in terms of $\mathcal{Y}_{ijk}$ as
\beq
{\sf c}_{123}=\frac{g^2}{2}\frac{x_{12}^2+x_{23}^2-2x_{13}^2}{\pi^2}\mathcal{Y}_{123}\period
\eeq
Using \eqref{eq:explicitconformal}, one can easily verify that the corner contribution satisfies the following useful identities
\beq\label{eq:cyclicity}
{\sf c}_{123}={\sf c}_{321}\comma\qquad {\sf c}_{123}+{\sf c}_{231}+{\sf c}_{312}=0\comma
\eeq
which will be important in the subsequent analysis. Another important feature is that the function ${\sf C}_{[12][34]}$ can be expressed in terms of ${\sf c}_{123}$,
\beq
{\sf C}_{[12][34]}=\frac{\bar{d}_{12}\bar{d}_{34}}{3}\left[({\sf c}_{312}-{\sf c}_{123})-({\sf c}_{412}-{\sf c}_{124})+({\sf c}_{134}-{\sf c}_{341})-({\sf c}_{234}-{\sf c}_{243})\right]\period
\eeq
In the computation below, we will also need the coincident limit of the corner contribution, which reads
\beq
\begin{aligned}
{\sf c}_{jjk}=-g^2\left(1+\log \frac{|x_{jk}|}{\epsilon}\right)\comma\qquad
{\sf c}_{jkj}=2g^2\left(1+\log \frac{|x_{jk}|}{\epsilon}\right)\period\\
\end{aligned}
\eeq
\paragraph{Boundary Diagrams}
Let us next discuss the diagrams that contain the contraction with the Wilson loop. The most basic diagram of this type at one loop involves the scalar-scalar-gluon cubic vertex in which the two scalars come from the operator insertions while the gluon gets contracted with the Wilson loop.
\begin{figure}[t]
\centering
\includegraphics[clip,height=4cm]{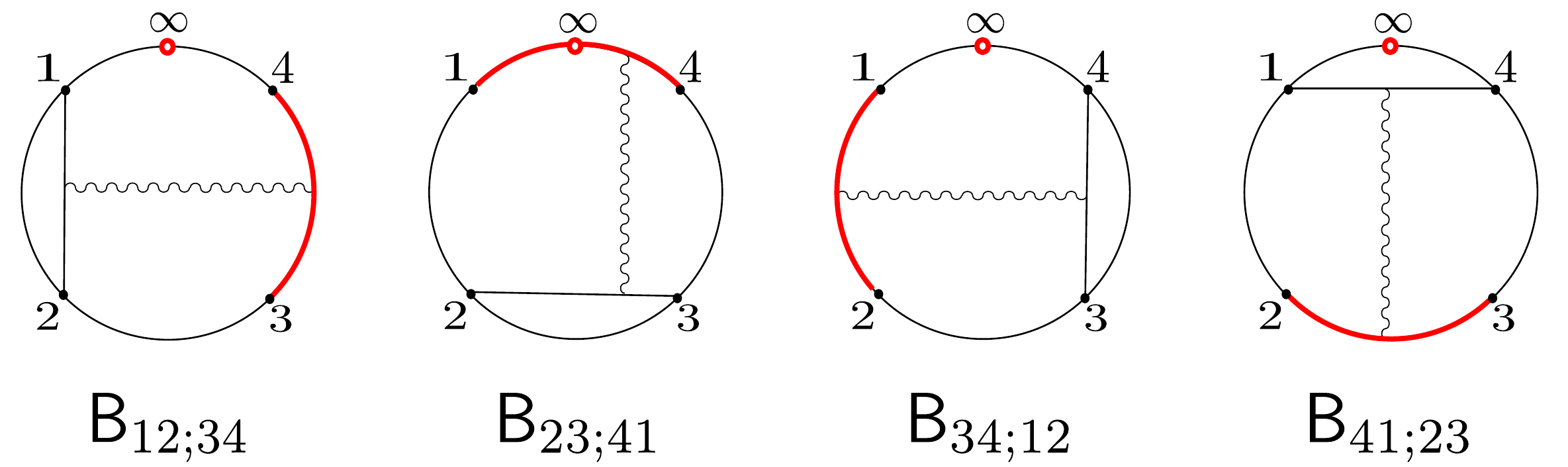}
\caption{Boundary diagrams relevant for the four-point functions: The wavy line denotes the gluon propagator and the red curve denotes the range of integration. Note that, although the Wilson loop is depicted as a circle in the figure, the computation was done for the straight-line Wilson loop. In order to keep in mind this point, we depicted a point at infinity in the figures and denoted it by the red dot.}
\label{fig:bdy1}
\end{figure}

To compute such a contribution, we simply need to bring down the cubic vertex from the action (see Appendix \ref{ap:action} for our convention of the action) and compute the following correlator using the tree-level Wick contraction,
\beq
\begin{aligned}
{\sf B}_{ij;kl}=&\Big<i\int^{x_l}_{x_k}d\tau \,\,{\rm Tr}{\rm P}\left[(Y_i\cdot \Phi)(x_i)(Y_j\cdot \Phi)(x_j)\mathcal{A}(\tau)\right] \left(\frac{2i}{g_{\rm YM}^2}\int d^4 x{\rm Tr}\left[ \del_{\mu}\Phi (x)[A^{\mu}(x),\Phi (x)]\right]\right)\Big>\nonumber
\end{aligned}
\eeq
where $\mathcal{A}(\tau)$ is the gauge connection along the Wilson loop, $\mathcal{A}(\tau)\equiv A_{\mu}\frac{dx^{\mu}}{d\tau}$. The result is given in terms of $\mathcal{Y}_{ijk}$ as
\beq
{\sf B}_{ij;kl}=-\frac{g^{4}(Y_i\cdot Y_j)}{\pi^2}\int_{x_k}^{x_l} d\tau \left(\del_i \mathcal{Y}_{ij\tau}-\del_{j}\mathcal{Y}_{ij\tau}\right)  \period
\eeq
This integral can be evaluated using the explicit form of $\mathcal{Y}_{123}$, \eqref{eq:explicitconformal}. The results read\fn{One potentially subtle point is that the result of the integral contains the (di-)logarithms, and depending on the ordering of $x_i$-$x_l$, one has to choose different branches of the logarithms. There are two strategies to overcome this subtlety: The one is to evaluate each case separately. The other is to evaluate a diagram for one particular ordering and then analytically continue the result to obtain a different ordering. The result of the analytic continuation is ambiguous due to the choice of the branches of the logarithms, but one can fix the ambiguity by requiring the reality of the final answer. We tried both approaches and got the same answer.} (see also figure \ref{fig:bdy1})
 \beq
 \begin{aligned}
 \frac{{\sf B}_{12;34}}{\bar{d}_{12}}=&4g^2 \left[ L_{R}\left(\frac{x_{21}}{x_{41}}\right)- L_{R}\left(\frac{x_{21}}{x_{31}}\right)\right]-\frac{{\sf c}_{123}+{\sf c}_{412}-{\sf c}_{124}-{\sf c}_{312}}{3}\comma\\
 \frac{{\sf B}_{34;12}}{\bar{d}_{34}}=&4g^2 \left[- L_{R}\left(\frac{x_{43}}{x_{42}}\right)+ L_{R}\left(\frac{x_{43}}{x_{41}}\right)\right]-\frac{{\sf c}_{341}+{\sf c}_{234}-{\sf c}_{342}-{\sf c}_{134}}{3}\comma\\
 \frac{{\sf B}_{23;41}}{\bar{d}_{23}}=&4g^2 \left[- L_{R}\left(\frac{x_{32}}{x_{31}}\right)- L_{R}\left(\frac{x_{32}}{x_{42}}\right)\right]-\frac{{\sf c}_{234}+{\sf c}_{123}-{\sf c}_{231}-{\sf c}_{423}}{3}\comma\\
\frac{{\sf B}_{41;23}}{\bar{d}_{14}}=&4g^2 \left[-L_{R}\left(\frac{x_{43}}{x_{41}}\right)+ L_{R}\left(\frac{x_{42}}{x_{41}}\right)\right]-\frac{{\sf c}_{412}+{\sf c}_{341}-{\sf c}_{413}-{\sf c}_{241}}{3}\comma
 \end{aligned}
 \eeq
 where $L_{R} (x)$ is the Rogers dilogarithm defined by
 \beq
 L_{R} (x)={\rm Li}_2 (x)+\frac{1}{2}\log x \,\log (1-x)\period
 \eeq
 The two important properties satisfied by the Rogers dilogarithm are
 \beq\label{eq:rogersid}
 \begin{aligned}
 &L_{R}(x)+L_{R}(1-x)=\frac{\pi^2}{6}\comma\\
 &L_{R}(x)+L_{R}(y)-L_{R}\left(\frac{x(1-y)}{1-x y}\right)-L_{R}\left(\frac{y(1-x)}{1-x y}\right)-L_{R}\left(x y\right)=0\period
  \end{aligned}
 \eeq
 
\begin{figure}[t]
\centering
\includegraphics[clip,height=3cm]{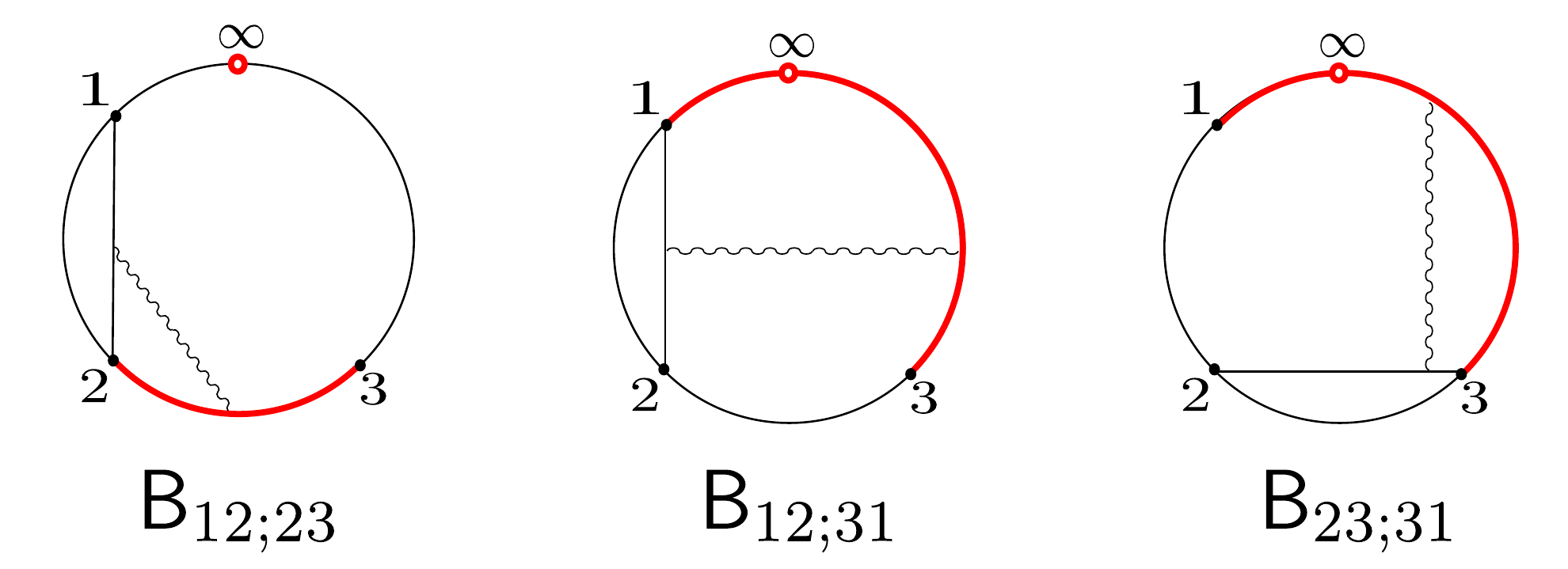}\\
\includegraphics[clip,height=3cm]{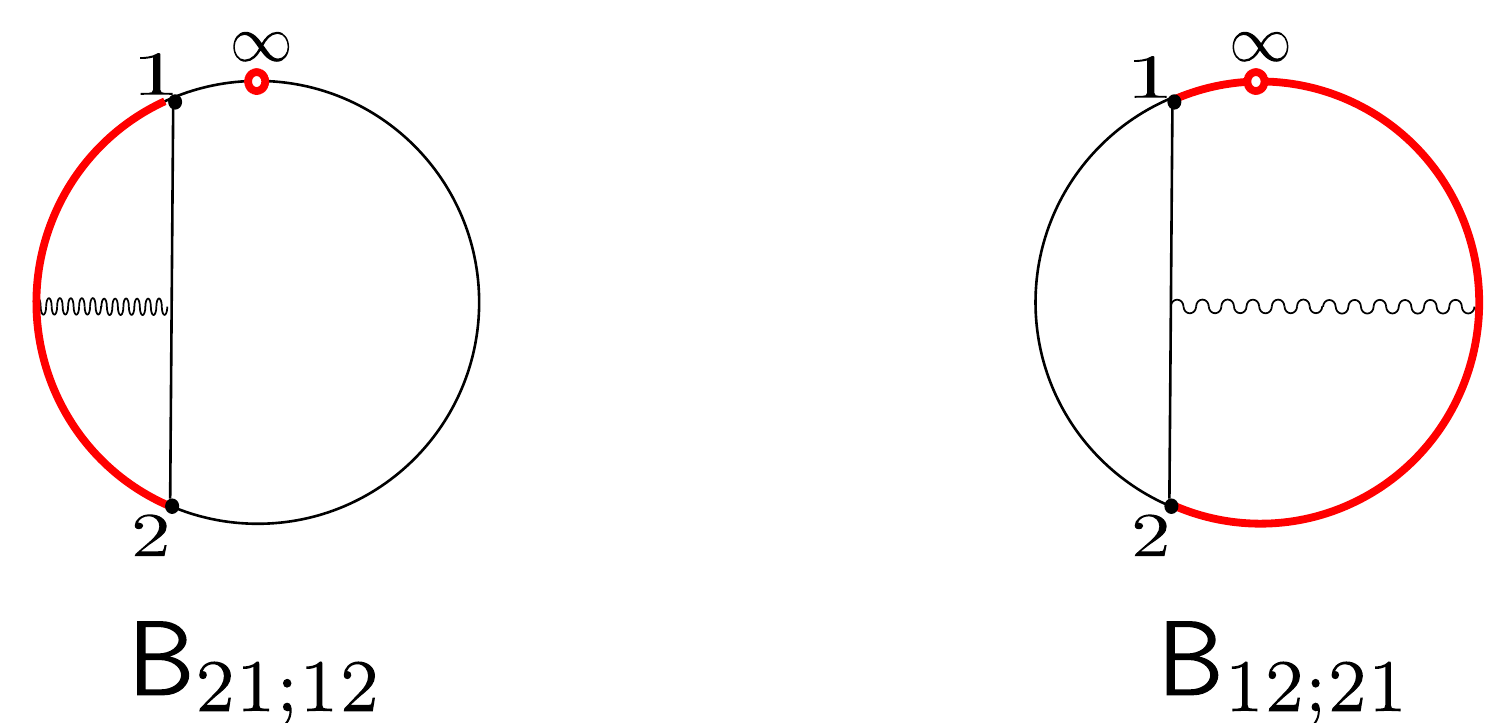}
\caption{Boundary diagrams relevant for the three-point functions (the first row) and the two-point functions (the second row). All these diagrams can be obtained by taking the limit of the figure \ref{fig:bdy1}.}
\label{fig:bdy2}
\end{figure}

 All the other relevant diagrams can be obtained by taking the coincident limit of some of the points. For instance, the diagrams that appear in the computation of the three-point functions are given by
 \beq
 \begin{aligned}
 \frac{{\sf B}_{12;23}}{\bar{d}_{12}}=&-4g^2 L_{R}\left(\frac{x_{32}}{x_{31}}\right)+g^2\left(1+\log \frac{|x_{12}|}{\epsilon}\right)+\frac{{\sf c}_{123}-{\sf c}_{312}}{3}\comma\\
 \frac{{\sf B}_{12;31}}{\bar{d}_{12}}=&-4g^2 \left[\frac{\pi^2}{6}+L_{R}\left(\frac{x_{21}}{x_{31}}\right)\right]+g^2\left(1+\log \frac{|x_{12}|}{\epsilon}\right)+\frac{{\sf c}_{312}-{\sf c}_{123}}{3}\comma\\
 \frac{{\sf B}_{23;31}}{\bar{d}_{12}}=&-4g^2 \left[\frac{\pi^2}{6}+L_{R}\left(\frac{x_{32}}{x_{31}}\right)\right]+g^2\left(1+\log \frac{|x_{23}|}{\epsilon}\right)+\frac{{\sf c}_{231}-{\sf c}_{123}}{3}\comma
 \end{aligned}
 \eeq
 For an explicit diagrammatic representation of each diagram, see figure \ref{fig:bdy2}.
 Similarly the diagrams that are relevant for the two-point functions can be computed as follows:
 \beq\label{eq:2ptbdy}
 \begin{aligned}
 \frac{{\sf B}_{12;21}}{\bar{d}_{12}}=&2g^{2}\left(1+\log \frac{|x_{12}|}{\epsilon}\right)-\frac{4\pi^2 g^2}{3}\comma\\
 \frac{{\sf B}_{21;12}}{\bar{d}_{12}}=&2g^{2}\left(1+\log \frac{|x_{12}|}{\epsilon}\right)+\frac{2\pi^2 g^2}{3}\period
 \end{aligned}
 \eeq
 Note that these two diagrams would look similar if we were considering the correlator on a circular Wilson loop. Here the difference comes from the fact that one of them includes the integration around the spatial infinity while the other does not.

\subsection{Two- and three-point functions}
Let us now compute the two- and the three-point functions using the results in the previous subsection.
\paragraph{Two-point functions}
At one loop, the two-point functions of length $L$ operators get corrections from the following diagrams:
\beq
\begin{aligned}
(L-1)\times {\sf G}_{12|21}\comma\qquad (L-1)\times {\sf Q}_{1221}\comma\qquad L\times {\sf S}_{12}\comma\qquad 1\times {\sf B}_{12;21}\comma\qquad 1\times {\sf B}_{21;12}\period
\end{aligned}
\eeq
One can simplify the computation using the identity,
\beq\label{eq:simpleid}
{\sf G}_{12|21}+{\sf Q}_{1221}+{\sf S}_{12}\bar{d}_{12}=0\comma
\eeq
which can be verified using the results in the previous subsection. Thus, we obtain
\beq\label{eq:final2pt}
\llangle \mathcal{O}_L (x_1) \mathcal{O}_{L}(x_2)\rrangle\big|_{O(g^2)}= \left({\sf S}_{12}+{\sf B}_{12;21}+{\sf B}_{21;12}\right)(\bar{d}_{12})^{L-1}=-\frac{2\pi^2g^2}{3}(2g^2)^{L}d_{12}^{L}\period
\eeq
It is perhaps worth emphasizing that, unlike the single-trace operators, the one-loop correction to the two-point function on the Wilson loop is non-zero. In other words, the normalization constant $n_L$ is nontrivial and is given by
\beq
n_{L}=(2g^2)^{L}\left[1-\frac{2\pi^2g^2}{3} +O(g^{4})\right]\period
\eeq
Note that this result is in perfect agreement with the result computed from localization in \cite{Giombi:2018qox}.
\begin{figure}[t]
\centering
\includegraphics[clip,height=4cm]{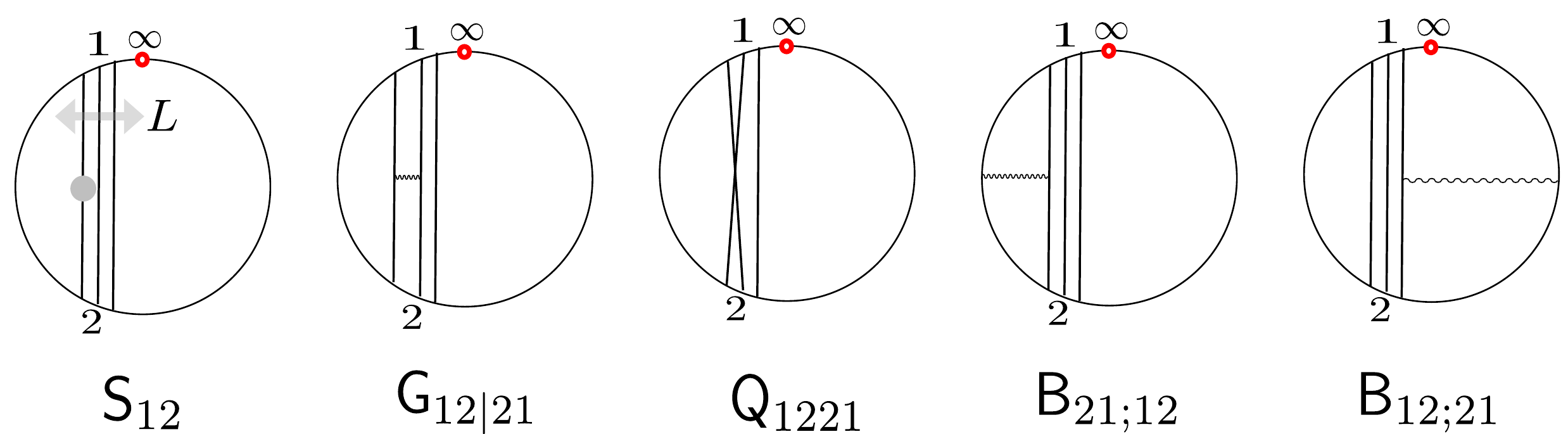}
\caption{The one-loop corrections to the two-point functions. Most of the diagrams cancel out owing to \eqref{eq:simpleid}, and the final result is given by \eqref{eq:final2pt}.}
\label{fig:twopt}
\end{figure}
\paragraph{Three-point functions}
Let us next discuss the three-point functions. At tree level, the three-point function is given simply by a planar Wick contraction, and the number of propagators between a pair of operators is determined completely by the lengths of the operators: For instance, the number of propagators between $\mathcal{O}_{L_i}$ and $\mathcal{O}_{L_j}$ is given by
\beq
\ell_{ij}\equiv \frac{L_i+L_j-L_k}{2}\comma
\eeq 
where $i,j$ and $k$ are a cyclic permutation of the indices $1,2$ and $3$. 

\begin{figure}[t]
\centering
\includegraphics[clip,height=10cm]{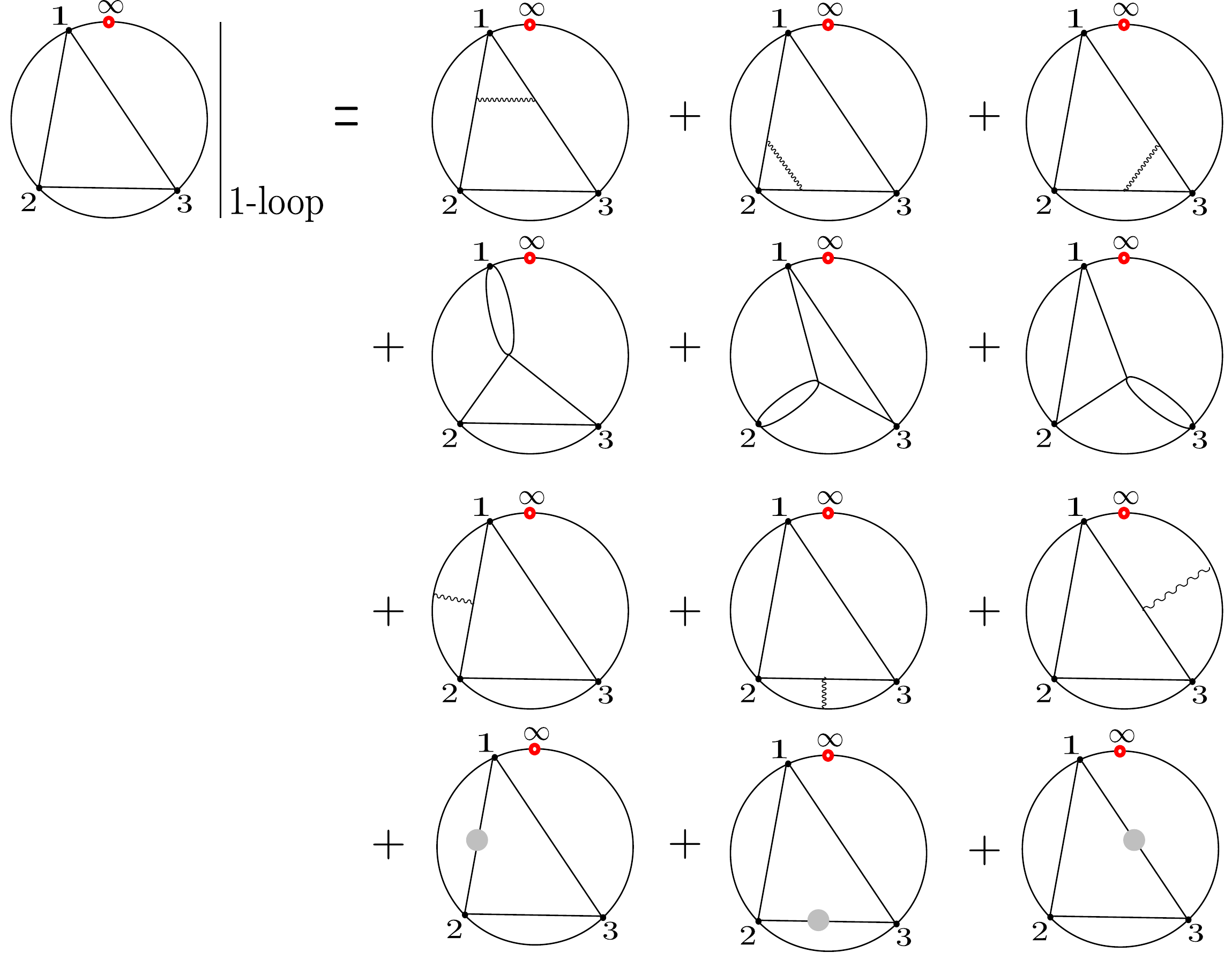}
\caption{The one-loop corrections to the three-point function with $L_1=L_2=L_3=2$. The first line and the second line denote the contributions from ${\sf G}_{ij|jk}$ and ${\sf Q}_{ijjk}$ respectively while the third line denotes the contribution from the boundary diagrams ${\sf B}_{jk;kj}$. The last line denotes the contribution from the self-energy diagrams.
The diagrams add up to \eqref{eq:sumisvanishing}.}
\label{fig:threept1}
\end{figure}

It turns out that the one-loop correction depends on whether $\ell_{ij}$'s are nonzero or not. If all $\ell_{ij}$'s are nonzero, we have the following one-loop diagrams (see figure \ref{fig:threept1}):
\beq\label{eq:listforthree1}
\begin{aligned}
&\sum_{i<j}(\ell_{ij}-1) \times \left[{\sf G}_{ij|ji}+{\sf Q}_{ijji}\right]\comma\qquad \sum_{i<j}\ell_{ij}\times {\sf S}_{ij}\comma\\&\sum_{\{i,j,k\}=\{1,2,3\}}{\sf G}_{ij|jk}+{\sf Q}_{ijjk}\comma\qquad {\sf B}_{21;12}+{\sf B}_{32;23}+{\sf B}_{13;31}\comma
\end{aligned}
\eeq
where the third term is the sum over the cyclic permutation of $\{1,2,3\}$.
Summing them up and using the identity \eqref{eq:simpleid}, we obtain 
\beq
\begin{aligned}
\bar{d}_{12}^{\ell_{12}}\bar{d}_{23}^{\ell_{23}}\bar{d}_{31}^{\ell_{31}}\left[\sum_{\{i,j,k\}=\{1,2,3\}}\frac{{\sf S}_{ij}+2 {\sf B}_{ji;ij}}{2 \bar{d}_{ij}}+\underbrace{\frac{1}{\bar{d}_{ij}\bar{d}_{jk}}\left(\frac{{\sf S}_{ij}\bar{d}_{jk}+\bar{d}_{ij}{\sf S}_{jk}}{4}+{\sf G}_{ij|jk}+{\sf Q}_{ijjk}\right)}_{{\sf c}_{ijk}}\right]\period\nonumber
\end{aligned}
\eeq
Note that we split the contributions from ${\sf S}_{ij}$ into two pieces and reorganized the sum.
It turned out that the second term in the summand coincides with the corner contribution ${\sf c}_{ijk}$. One can then show that their contribution vanishes owing to the cyclicity property of ${\sf c}_{ijk}$ \eqref{eq:cyclicity}. Furthermore, the first term in the summand can be computed using  \eqref{eq:selfenergy} and \eqref{eq:2ptbdy} as
\beq
\frac{{\sf S}_{12}}{2}+{\sf B}_{21;12}=\frac{2\pi^2g^2}{3}\bar{d}_{12}\comma\quad
\frac{{\sf S}_{23}}{2}+{\sf B}_{32;23}=\frac{2\pi^2g^2}{3}\bar{d}_{23}\comma\quad
\frac{{\sf S}_{31}}{2}+{\sf B}_{13;31}=-\frac{4\pi^2g^2}{3}\bar{d}_{31}\period
\eeq
The asymmetry between the first two and the last terms come from the fact that ${\sf B}_{13;31}$ includes the integration at infinity while the other integrals do not.
The sum of all these terms turned out to vanish and we get
\beq\label{eq:sumisvanishing}
\left.\llangle \mathcal{O}_{L_1}(x_1)\mathcal{O}_{L_2}(x_2)\mathcal{O}_{L_3}(x_3)\rrangle\right|_{O(g^2)} =0\qquad (\ell_{ij}\neq 0)\period
\eeq

\begin{figure}[t]
\centering
\includegraphics[clip,height=5cm]{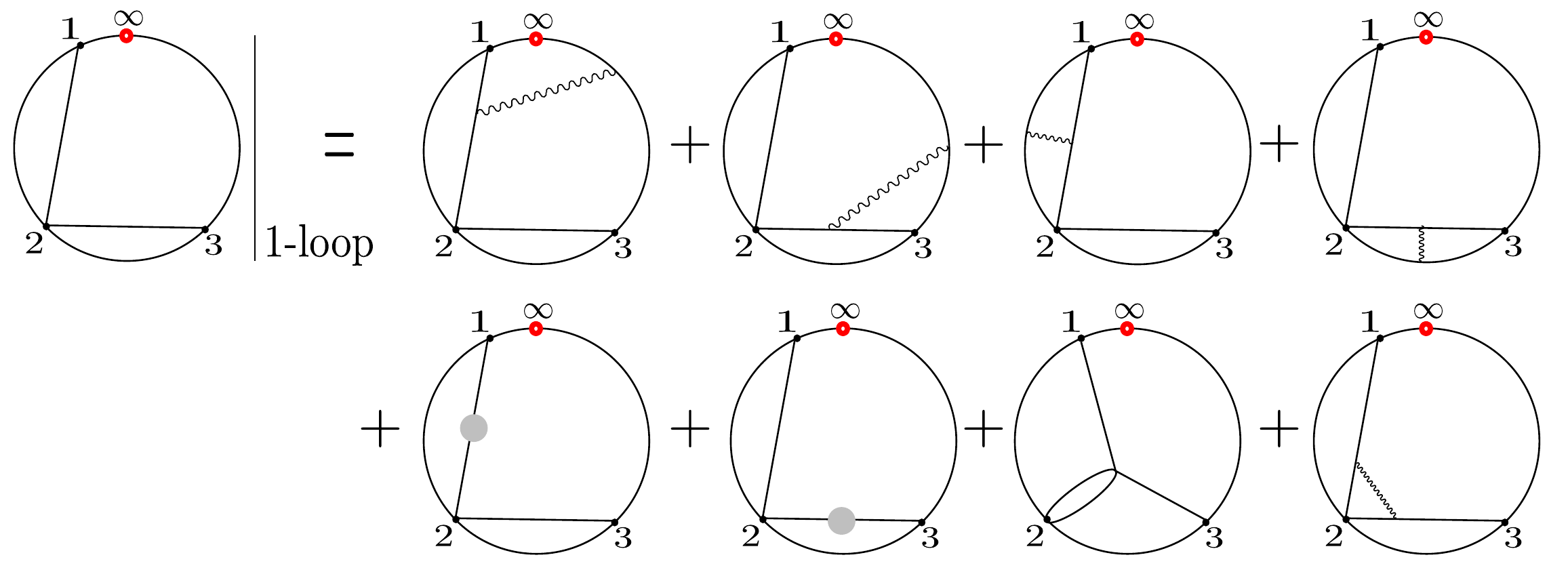}
\caption{The one-loop corrections to the three-point function with $L_1=L_2=2, L_3=1$.
The final result is given by \eqref{eq:sumisnot}.}
\label{fig:threept2}
\end{figure}

Let us next study the case where one of $\ell_{ij}$'s, say $\ell_{31}$, is zero. In this case, we lose the diagrams 
\beq
{\sf G}_{31|13}+{\sf Q}_{3113}\comma\quad {\sf S}_{31}\comma\quad {\sf G}_{23|31}+{\sf Q}_{2331}\comma\quad {\sf G}_{31|12}+{\sf Q}_{3112}\comma\quad {\sf B}_{13;31}\period
\eeq
 On the other hand, there are also new diagrams which are
\beq
\begin{aligned}
{\sf B}_{12;31}\comma\qquad {\sf B}_{23;31}\period
\end{aligned}
\eeq
Summing all the contributions (see also figure \ref{fig:threept2}), we obtain
\begin{align}\label{eq:sumisnot}
&{\sf c}_{123}d_{12}^{\ell_{12}}d_{23}^{\ell_{23}}+
d_{12}^{\ell_{12}-1}d_{23}^{\ell_{23}-1}\left(3\frac{{\sf S}_{12}d_{23}+d_{12}{\sf S}_{23}}{4}+({\sf B}_{12;31}+{\sf B}_{21;12})d_{23}+d_{12}({\sf B}_{23;31}+{\sf B}_{32;23})\right)\nonumber\\
&=-\frac{2\pi^2 g^2}{3}\period
\end{align}

Combining the two results, we obtain the following expression for the general three-point function,
\beq\label{eq:3ptfinaloneloop1}
\begin{aligned}
&\left.\llangle \mathcal{O}_{L_1}(x_1)\mathcal{O}_{L_2}(x_2)\mathcal{O}_{L_3}(x_3)\rrangle\right|_{O(g^2)} \\
&=-\bar{d}_{12}^{\ell_{12}}\bar{d}_{23}^{\ell_{23}}\bar{d}_{31}^{\ell_{31}}\frac{2\pi^2g^2 (\delta_{L_1+L_2,L_3}+\delta_{L_2+L_3,L_1}+\delta_{L_3+L_1,L_2})}{3}\period
\end{aligned}
\eeq
Dividing them with the normalization $n_{L_i}$, we can compute the structure constant as
\beq\label{eq:3ptfinal}
\begin{aligned}
&C_{L_1,L_2,L_3}\left(=\frac{\llangle \mathcal{O}_{L_1}(x_1)\mathcal{O}_{L_2}(x_2)\mathcal{O}_{L_3}(x_3)\rrangle}{\sqrt{n_{L_1}n_{L_2}n_{L_3}}}\right)\\
&=1+\pi^{2}g^{2}\left[1-\frac{2}{3}(\delta_{L_1+L_2,L_3}+\delta_{L_2+L_3,L_1}+\delta_{L_3+L_1,L_2})\right]+O(g^{4})\period
\end{aligned}
\eeq
These results coincide with the localization computation in \cite{Giombi:2018qox}.

\section{Pertubation II: Four-Point Functions}
\subsection{General structure of tree-level contractions}
Before discussing the one-loop correction, let us first take a look at the general structure of the tree-level contractions. 

\begin{figure}[t]
\centering
\includegraphics[clip,height=5cm]{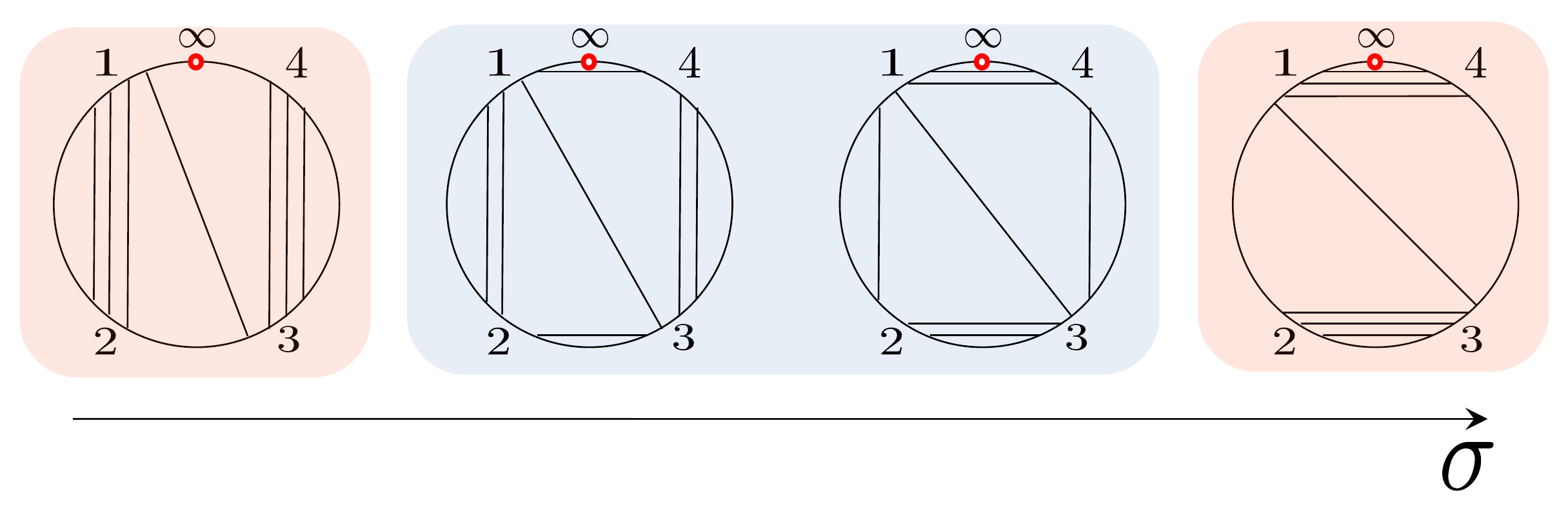}
\caption{The one-dimensional ``moduli space'' of the diagrams for $L_1=L_3=4, L_2=L_4=3$. The ones drawn on the red background correspond to the boundaries of the moduli space while the ones drawn on the blue background correspond to the bulk of the moduli space.}
\label{fig:moduli}
\end{figure}
Unlike the case for the two- and the three-point functions, there are multiple ways to perform the planar Wick contractions of four operators, and the tree-level answer  is given by a sum of such contractions. Each diagram is specified by the numbers of propagators between two operators $\ell_{ij}$, which fall into two different classes; namely the ``neighboring contractions''
\beq
\ell_{12}\comma\quad \ell_{23}\comma\quad \ell_{34}\comma\quad \ell_{41}\comma
\eeq
and the ``diagonal contractions''
\beq
\ell_{13}\comma\quad \ell_{24}\period
\eeq
Owing to the planarity, only one of the diagonal contractions can be nonzero, and they are determined by the lengths of four operators\fn{This can be shown easily by writing the length of the operators in terms of $\ell_{ij}$, for instance $L_1=\ell_{12}+\ell_{13}+\ell_{41}$, and taking an appropriate linear combination.}:
\beq
\begin{aligned}
L_1+L_3\geq L_2+L_4: & \quad \ell_{13}=\frac{(L_1+L_3)-(L_2+L_4)}{2}\comma\quad &&\ell_{24}=0\comma\\
L_1+L_3\leq L_2+L_4: & \quad \ell_{24}=\frac{(L_2+L_4)-(L_1+L_3)}{2}\comma\quad &&\ell_{13}=0\period
\end{aligned}
\eeq
Note that, when $L_1+L_3=L_2+L_4=0$, there are no diagonal contractions, $\ell_{13}=\ell_{24}=0$.
On the other hand, the lengths of the neighboring contractions are not fixed and there is a one-dimensional ``moduli space''\fn{One can view this as a discretized version of the one-dimensional of a disk with four marked points.} of diagrams which is parametrized by the value of $\sigma\equiv (\ell_{14}+\ell_{23})/2$ (see figure \ref{fig:moduli} for an example). The sum over the moduli space is given by a geometric series with respect to the cross ratio in the following way:
\beq
\begin{aligned}
\left.\llangle\mathcal{O}_{L_1}\mathcal{O}_{L_2}\mathcal{O}_{L_3}\mathcal{O}_{L_4} \rrangle\right|_{\text{tree}}&=(\text{Diagram with $\sigma_{\rm min}$})\times \sum_{\sigma=\sigma_{\rm min}}^{\sigma_{\rm max}}\left[\frac{\bar{d}_{14}\bar{d}_{23}}{\bar{d}_{12}\bar{d}_{34}}\right]^{\sigma-\sigma_{\rm min}}\\
&=(\text{Diagram with $\sigma_{\rm min}$})\times\sum_{\sigma=\sigma_{\rm min}}^{\sigma_{\rm max}}\left[\frac{\chi^{2}}{(1-\chi)^2}\frac{(1-\alpha)(1-\bar{\alpha})}{\alpha\bar{\alpha}}\right]^{\sigma-\sigma_{\rm min}}\comma
\end{aligned}
\eeq
where $\sigma_{\rm min}$ and $\sigma_{\rm max}$ are the minimal and the maximal values $\sigma$ can take. It is also possible to write down explicitly the contribution from the diagram with $\sigma_{\rm min}$. We then get
\beq\label{eq:treelevelfinal}
\left.\llangle\mathcal{O}_{L_1}\mathcal{O}_{L_2}\mathcal{O}_{L_3}\mathcal{O}_{L_4} \rrangle\right|_{\text{tree}}={\sf D}_{\rm left}\times\sum_{n=0}^{n_{\rm max}}\left[\frac{\chi^{2}}{(1-\chi)^2}\frac{(1-\alpha)(1-\bar{\alpha})}{\alpha\bar{\alpha}}\right]^{n}\comma
\eeq
where ${\sf D}_{\rm left}$ is the contribution from the diagram at the left edge of the moduli space,
\beq
{\sf D}_{\rm left}=\bar{d}_{12}^{\ell^{\rm max}_{12}}\bar{d}_{34}^{\ell^{\rm max}_{34}}\bar{d}_{41}^{\ell^{\rm min}_{41}}\bar{d}_{23}^{\ell^{\rm min}_{23}}\bar{d}_{13}^{\ell_{13}}\bar{d}_{24}^{\ell_{24}}\comma
\eeq
with
\beq
n_{\rm max}\equiv {\rm min}\left[L_1-\ell_{13},L_2-\ell_{24},L_3-\ell_{13},L_4-\ell_{24}\right]\comma
\eeq
and
\beq
\begin{aligned}
\ell_{13}&={\rm max}\left[\frac{L_1-L_2+L_3-L_4}{2},0\right]\comma\quad &&\ell_{24}={\rm max}\left[\frac{-L_1+L_2-L_3+L_4}{2},0\right]\comma\\
\ell_{12}^{\rm max}&={\rm min}\left[L_1-\ell_{13},L_2-\ell_{24}\right]\comma\quad 
&&\ell_{34}^{\rm max}={\rm min}\left[L_3-\ell_{13},L_4-\ell_{24}\right]\comma \\
\ell_{41}^{\rm min}&={\rm max}\left[L_1-L_2-\ell_{13}+\ell_{24},0\right]\comma\quad &&\ell_{23}^{\rm min}={\rm max}\left[-L_1+L_2+\ell_{13}-\ell_{24},0\right]\period
\end{aligned}
\eeq
Similarly, the contribution from the diagram at the right edge of the moduli space reads
\beq
{\sf D}_{\rm right}={\sf D}_{\rm left}\left[\frac{\chi^{2}}{(1-\chi)^2}\frac{(1-\alpha)(1-\bar{\alpha})}{\alpha\bar{\alpha}}\right]^{n_{\rm max}}=\bar{d}_{12}^{\ell^{\rm min}_{12}}\bar{d}_{34}^{\ell^{\rm min}_{34}}\bar{d}_{41}^{\ell^{\rm max}_{41}}\bar{d}_{23}^{\ell^{\rm max}_{23}}\bar{d}_{13}^{\ell_{13}}\bar{d}_{24}^{\ell_{24}}\comma
\eeq
with
\beq
\begin{aligned}
\ell_{41}^{\rm max}&={\rm min}\left[L_1-\ell_{13},L_4-\ell_{24}\right]\comma\quad 
&&\ell_{23}^{\rm max}={\rm min}\left[L_2-\ell_{24},L_3-\ell_{13}\right]\comma \\
\ell_{12}^{\rm min}&={\rm max}\left[L_1-L_4-\ell_{13}+\ell_{24},0\right]\comma\quad &&\ell_{34}^{\rm min}={\rm max}\left[-L_1+L_4+\ell_{13}-\ell_{24},0\right]\period
\end{aligned}
\eeq

\begin{figure}[t]
\centering
\includegraphics[clip,height=8cm]{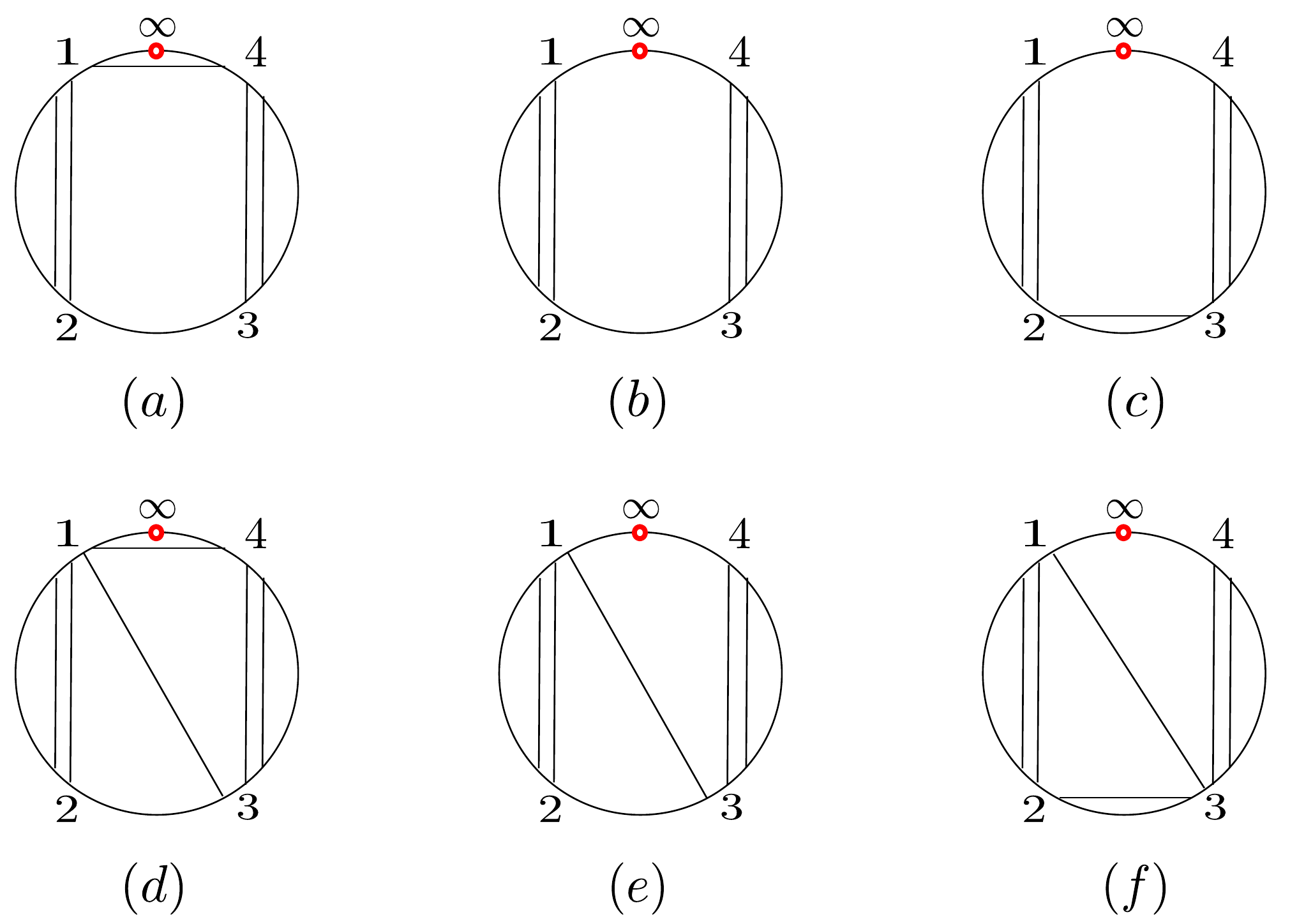}
\caption{Examples of the diagrams that live at the left boundary of the moduli space. $(a)$-$(c)$ are the ones without diagonal contractions while $(d)$-$(f)$ are the ones with diagonal contractions. Depending on the relative lengths of the operators, the diagrams have different topology at the boundary of the moduli space. }
\label{fig:various}
\end{figure}
In the bulk of this moduli space, the diagrams are always of the ``square shape''; namely all the neighboring contractions are nonzero (see figure \ref{fig:moduli}). At the boundary of the moduli space, however, the structure of the diagrams depends on the relative magnitude of the lengths $L_k$'s (see figure \ref{fig:various}).
\subsection{One-loop correction to the bulk diagrams}
\begin{figure}[t]
\centering
\includegraphics[clip,height=6cm]{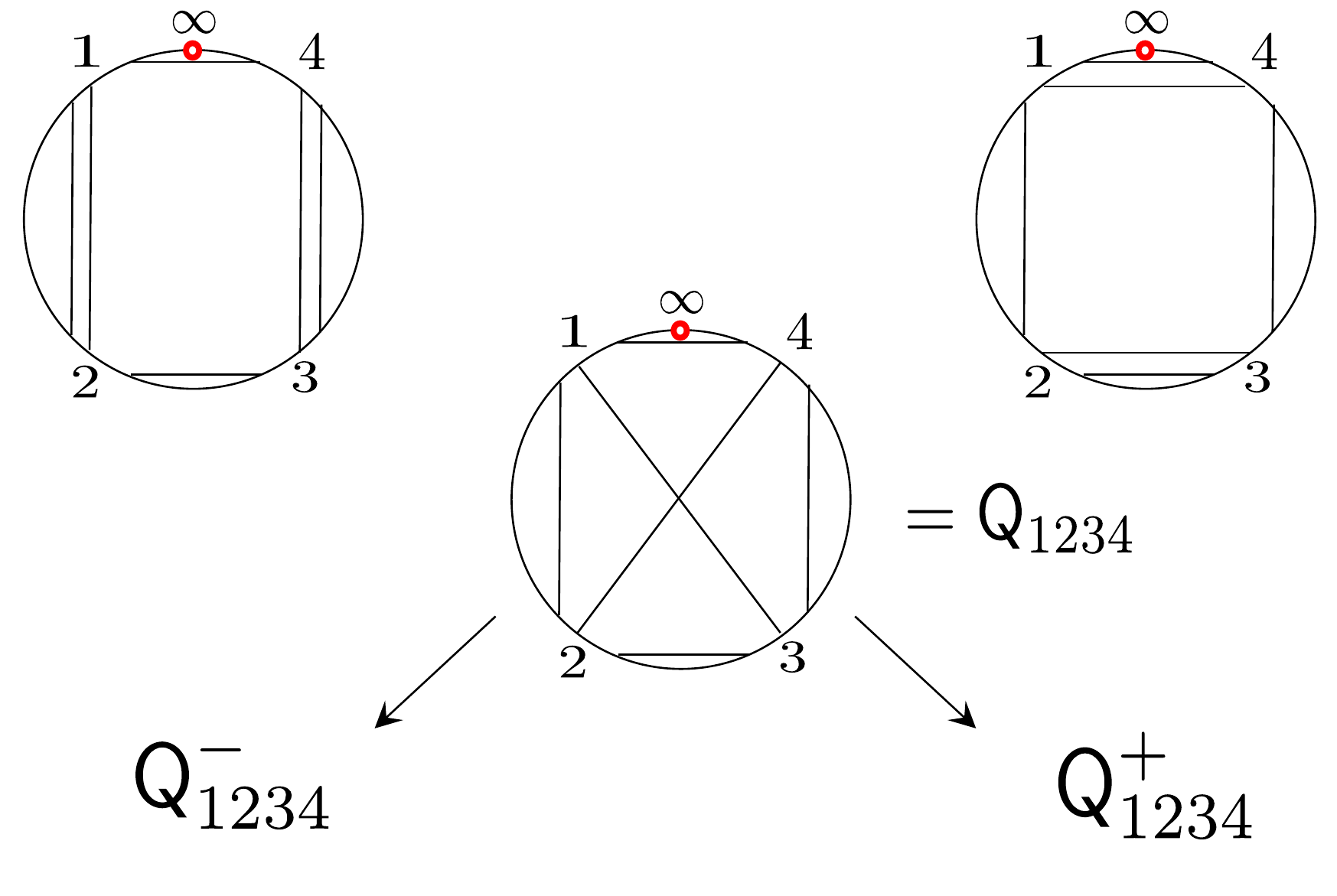}
\caption{The decomposition of the scalar quartic diagram. The scalar quartic diagram sits ``in the middle'' of two different tree-level graphs. By decomposing the scalar quartic diagram into two pieces \eqref{eq:howtodecompose}, one can associate each contribution with a tree-level diagram.}
\label{fig:decomposeQ}
\end{figure}

\begin{figure}[t]
\centering
\includegraphics[clip,height=3cm]{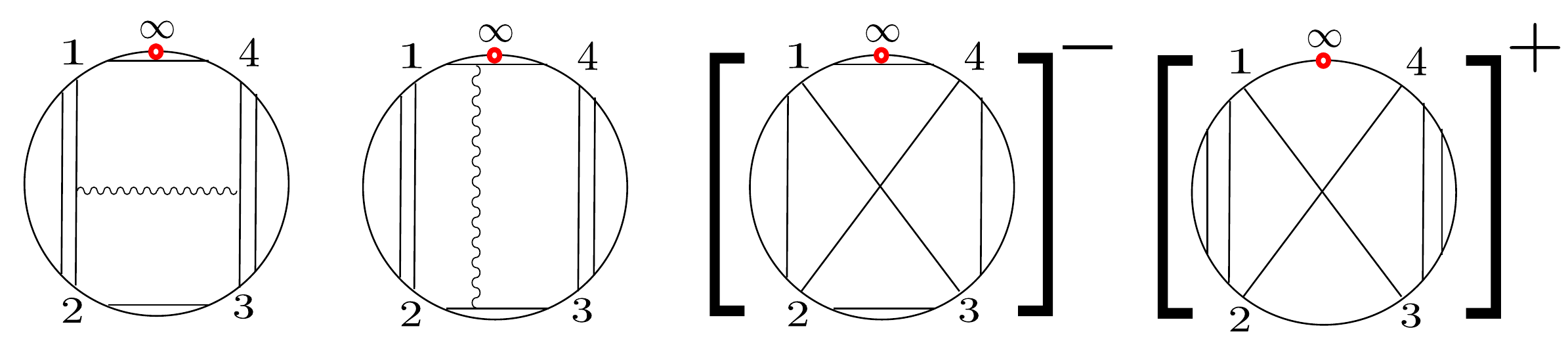}
\caption{Examples of one-loop diagrams for the four-point function without diagonal contractions $(\ell_{13}=\ell_{24}=0)$. The first two diagrams are the gluon-exchange diagrams while the third and the fourth diagrams are ${\sf Q}_{1234}^{-}$ and ${\sf Q}_{1234}^{+}$.}
\label{fig:4ptpic1}
\end{figure}

Let us now analyze the one-loop corrections to the bulk diagram, for which all neighboring contractions are nonzero.  
Most of the one-loop corrections can be obtained by dressing the tree-level diagrams with a gluon propagator. The only exception is the scalar quartic interaction ${\sf Q}_{1234}$.
In a sense, the scalar quartic diagram sits between two tree-level diagrams. This can be seen easily from the fact that the expression \eqref{eq:defGQ} contains several different contractions, $\bar{d}_{ij}\bar{d}_{kl}$. This small complication  can be overcome by decomposing ${\sf Q}_{1234}$ as follows and treating each term separately (see also figure \ref{fig:decomposeQ}):
\beq\label{eq:howtodecompose}
\begin{aligned}
&{\sf Q}_{1234}={\sf Q}_{1234}^{-}+{\sf Q}_{1234}^{+}\comma\\
&{\sf Q}_{1234}^{-}=\frac{g^2}{2}\Phi (\chi)\chi^2 \left(\frac{1}{\alpha}+\frac{1}{\bar{\alpha}}-1\right)\bar{d}_{12}\bar{d}_{34}\comma\\
&{\sf Q}_{1234}^+=\frac{g^2}{2}\Phi (\chi)(1-\chi)^2 \left(\frac{1}{1-\alpha}+\frac{1}{1-\bar{\alpha}}-1\right)\bar{d}_{23}\bar{d}_{14}\period
\end{aligned}
\eeq
In what follows, we will use this decomposition and associate all the one-loop corrections with the individual tree-level diagrams.
\paragraph{Diagrams without diagonal contraction} With this in mind, we now study the bulk diagram without diagonal contractions, namely $\ell_{13}=\ell_{24}=0$. The one-loop corrections are given by\fn{Here we use the mod $4$ identification of the subscripts, namely $\ell_{45}\equiv \ell_{41}$. } (see figure \ref{fig:4ptpic1})
\beq\label{eq:fourlist1}
\begin{aligned}
&\sum_{k}\ell_{k\,k+1} {\sf S}_{k\, k+1}\comma\quad  \sum_{k}(\ell_{k \,k+1}-1)({\sf G}_{k\, k+1|k+1\,k}+{\sf Q}_{k \,k+1\,k+1\,k})\comma \quad \sum_k {\sf B}_{k+1\,k;k\,k+1}\\
&\sum_{k}{\sf G}_{k\, k+1|k+1\,k+2}+{\sf Q}_{k \,k+1\,k+1\,k+2}\comma\quad {\sf G}_{12|34}\comma\quad {\sf G}_{23|41}\comma\quad {\sf Q}_{1234}^{+}\comma \quad {\sf Q}_{1234}^{-}\period
\end{aligned}
\eeq
Using \eqref{eq:simpleid}, the first two terms in \eqref{eq:fourlist1} can be simplified into
\beq
{\sf S}_{12}\comma\quad {\sf S}_{23} \comma\quad {\sf S}_{34} \comma\quad {\sf S}_{41}\period
\eeq
Combining them with the third and the fourth terms in \eqref{eq:fourlist1}, we get
\begin{align}\label{eq:firstfirstfour}
&\frac{{\sf S}_{12}+2{\sf B}_{21;12}}{2\bar{d}_{12}}=\frac{{\sf S}_{23}+2{\sf B}_{32;23}}{2\bar{d}_{23}}=\frac{{\sf S}_{34}+2{\sf B}_{43;34}}{2\bar{d}_{34}}=\frac{2\pi^2g^2}{3}\comma\quad
 \frac{{\sf S}_{41}+2{\sf B}_{14;41}}{2\bar{d}_{41}}=-\frac{4\pi^2g^2}{3}\comma\\
 &\sum_{k}{\sf G}_{k\,k+1|k+1\,k}+{\sf Q}_{k\,k+1\,k+1\,k}+\frac{{\sf S}_{k\,k+1}\bar{d}_{k+1\,k+2}+\bar{d}_{k\,k+1}{\sf S}_{k+1\,k+2}}{4}={\sf c}_{123}+{\sf c}_{234}+{\sf c}_{341}+{\sf c}_{412}\period\nonumber
\end{align}
 Finally, the last two terms in \eqref{eq:fourlist1} give
 \beq
 \begin{aligned}\label{eq:secondfirstfour}
 &\frac{{\sf G}_{12|34}+{\sf Q}_{1234}^{-}}{\bar{d}_{12}\bar{d}_{34}}=\frac{g^2}{2}\Phi (\chi)\left(\frac{\chi^2}{\alpha}+\frac{\chi^2}{\bar{\alpha}}-2\chi\right)+\frac{{\sf C}_{[12][34]}}{\bar{d}_{12}\bar{d}_{34}}\comma\\
 &\frac{{\sf G}_{23|41}+{\sf Q}_{1234}^{+}}{\bar{d}_{23}\bar{d}_{41}}=\frac{g^2}{2}\Phi (\chi)\left(\frac{(1-\chi)^2}{1-\alpha}+\frac{(1-\chi)^2}{1-\bar{\alpha}}-2(1-\chi)\right)+\frac{{\sf C}_{[23][41]}}{\bar{d}_{23}\bar{d}_{41}}\period
 \end{aligned}
 \eeq
 Summing \eqref{eq:firstfirstfour} and \eqref{eq:secondfirstfour} and using
 \beq
 {\sf C}_{[12][34]}+{\sf C}_{[23][41]}=-{\sf c}_{123}-{\sf c}_{234}-{\sf c}_{341}-{\sf c}_{412}\comma
 \eeq
 we finally obtain the following answer
 \beq
 \left.{\tt bulk}\right|_{\ell_{13}=\ell_{24}=0}=\bar{d}_{12}^{\ell_{12}}\bar{d}_{23}^{\ell_{23}}\bar{d}_{34}^{\ell_{34}}\bar{d}_{41}^{\ell_{41}}\left[m\left(\frac{\chi}{\chi-1}\right)+m\left(\frac{\chi-1}{\chi}\right)+\frac{2\pi^2 g^2}{3}\right]\period
 \eeq
 Here $m(\chi)$ is the function introduced in \cite{Fleury:2016ykk}, which reads\fn{Although we do not write it explicitly, the function $m(\chi)$ also depends on $\alpha$ and $\bar{\alpha}$, and they are also transformed in the same way when we perform the transformation for $\chi$. Namely
 \beq
 m\left(\frac{\chi}{\chi-1}\right)\equiv g^2\left[\frac{\chi}{\chi-1}-\frac{1}{2}\left(\frac{\alpha}{\alpha-1}+\frac{\bar{\alpha}}{\bar{\alpha}-1}\right)\right]\Phi \left(\frac{\chi}{\chi-1}\right)\period
 \eeq
 }
 \beq
 m(\chi) =g^2 \left(\chi-\frac{\alpha+\bar{\alpha}}{2}\right)\Phi (\chi)\period
 \eeq

\begin{figure}[t]
\centering
\includegraphics[clip,height=4cm]{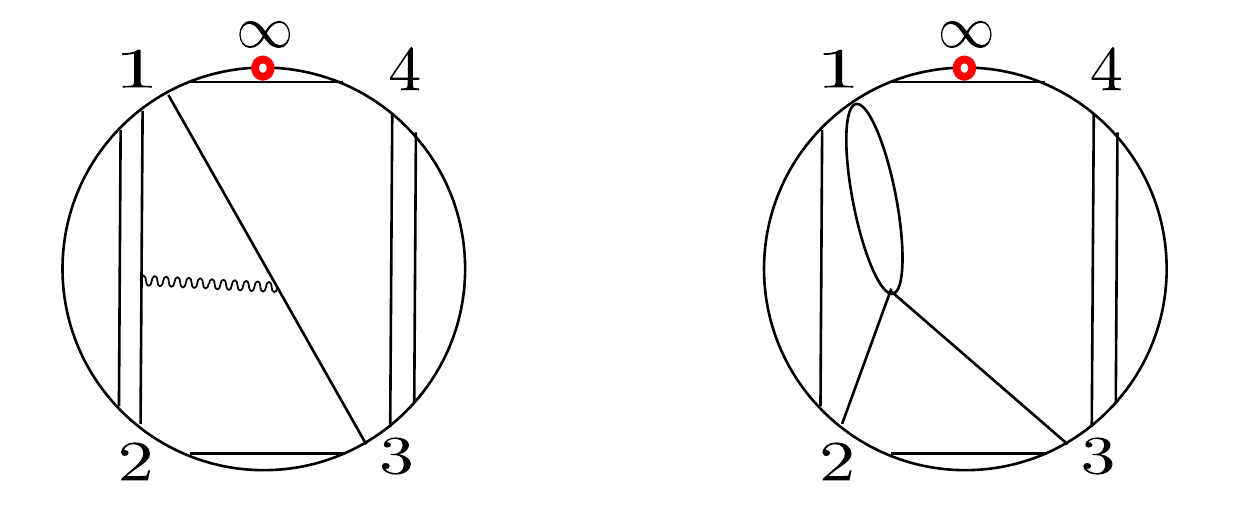}
\caption{Examples of one-loop diagrams for the four-point function with diagonal contractions $\ell_{13}\neq 0)$. All the diagrams are basically the same as the one for the three-point diagrams. Namely they can be expressed as the sum of corner contributions.}
\label{fig:4ptpic2}
\end{figure}

 \paragraph{Diagrams with diagonal contraction}
 Let us next analyze the bulk diagram with diagonal contractions. We focus on the case with $\ell_{13}\neq 0$ since the other case ($\ell_{24}\neq 0$) can be analyzed similarly. In the presence of the diagram contractions, one cannot draw a scalar quartic diagrams or a gluon exchange diagrams involving all four operators. Instead the one-loop corrections are given by (see also figure \ref{fig:4ptpic2})
 \begin{align}
&\sum_{k}\ell_{k\,k+1} {\sf S}_{k\, k+1}\comma\quad  \sum_{k}(\ell_{k \,k+1}-1)({\sf G}_{k\, k+1|k+1\,k}+{\sf Q}_{k \,k+1\,k+1\,k})\comma \quad \sum_k {\sf B}_{k+1\,k;k\,k+1}\label{eq:nodiagonalbulk}\\
&\ell_{13}{\sf S}_{13}\comma\quad(\ell_{13}-1)({\sf G}_{13|31}+{\sf Q}_{1331})\comma\quad \sum_{\{i,j,k\}=\{1,2,3\}}{\sf G}_{ij|jk}+{\sf Q}_{ijjk}\comma\quad \sum_{\{i,j,k\}=\{2,3,4\}}{\sf G}_{ij|jk}+{\sf Q}_{ijjk}\period\nonumber
\end{align}
By appropriately reorganizing the terms and using the identities \eqref{eq:simpleid} and \eqref{eq:cyclicity}, one can show that the sum of these diagrams reduce to
\beq
\frac{{\sf S}_{12}+2{\sf B}_{21;12}}{2\bar{d}_{12}}=\frac{{\sf S}_{23}+2{\sf B}_{32;23}}{2\bar{d}_{23}}=\frac{{\sf S}_{34}+2{\sf B}_{43;34}}{2\bar{d}_{34}}=\frac{2\pi^2g^2}{3}\period
\eeq
Summing them up, we get
\beq
 \left.{\tt bulk}\right|_{\ell_{13}\neq 0}=\bar{d}_{12}^{\ell_{12}}\bar{d}_{23}^{\ell_{23}}\bar{d}_{34}^{\ell_{34}}\bar{d}_{41}^{\ell_{41}}\bar{d}_{13}^{\ell_{13}}\frac{2\pi g^2}{3}\period
\eeq
Similarly, the case with $\ell_{24}\neq 0$ reads
\beq
\left.{\tt bulk}\right|_{\ell_{24}\neq 0}=\bar{d}_{12}^{\ell_{12}}\bar{d}_{23}^{\ell_{23}}\bar{d}_{34}^{\ell_{34}}\bar{d}_{41}^{\ell_{41}}\bar{d}_{24}^{\ell_{24}}\frac{2\pi g^2}{3}\period
\eeq

\subsection{One-loop correction to the boundary diagrams}
We now compute the corrections to the diagrams that live at the boundaries of the moduli space. In what follows, we focus on the diagrams at the left end of the moduli space, namely the diagrams with a minimal $\sigma\equiv (\ell_{14}+\ell_{23})/2$ since the diagrams at the other end can be obtained by the permutation of the operator labels (we will write down the explicit results in the next section).
\paragraph{Diagrams without diagonal contraction}
As in the previous subsection, let us first discuss the diagrams without diagonal contractions. In general, there are three different possibilities depending on the relative magnitude between $L_1+L_4$ and $L_2 +L_3$.

\begin{figure}[t]
\centering
\includegraphics[clip,height=4cm]{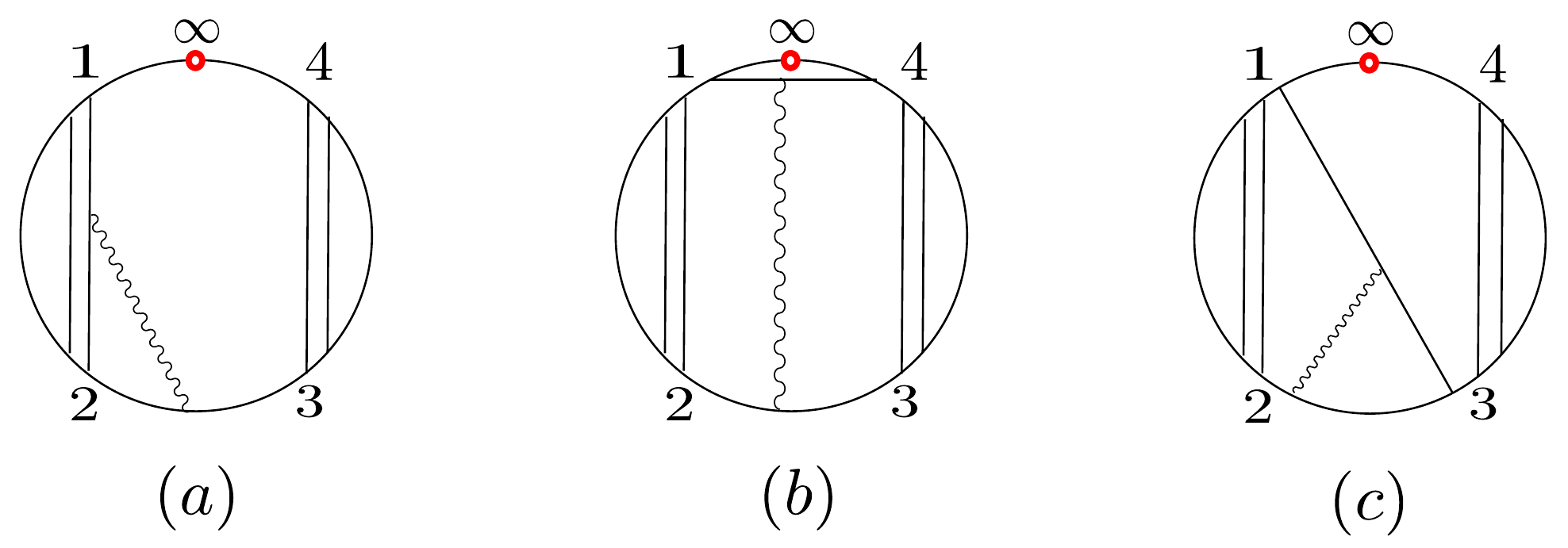}
\caption{Exmaples of the boundary diagrams at one loop. $(a)$ A boundary diagram without diagonal contractions for $\ell_{14}=\ell_{23}=0$. $(b)$ A boundary diagram without diagonal contractions for $\ell_{41}\neq 0$. $(c)$ A boundary diagram with diagonal contractions.}
\label{fig:sample}
\end{figure}
When $L_1+L_4$ and $L_2+L_3$ are equal, we have $\ell_{14}=\ell_{23}=0$, and the one-loop corrections are given by the following sets of diagrams (see also figure \ref{fig:sample}-$(a)$):
\beq
\begin{aligned}
&\ell_{12} {\sf S}_{12}\comma\quad (\ell_{12}-1)({\sf G}_{12|21}+{\sf Q}_{1221}) \comma\quad\ell_{34} {\sf S}_{34}\comma\quad (\ell_{34}-1)({\sf G}_{34|43}+{\sf Q}_{3443})\comma\\
&{\sf G}_{12|34}+{\sf Q}_{1234}^{-}\comma\quad {\sf B}_{21;12}\comma\quad {\sf B}_{43;34}\comma\quad {\sf B}_{12;23}\comma\quad {\sf B}_{12;41}\comma\quad {\sf B}_{34;23}\comma\quad {\sf B}_{34;41}\period
\end{aligned}
\eeq
As in the analysis in the previous subsection, the terms in the first line add up to ${\sf S}_{12}$ and ${\sf S}_{34}$. Combining them with the rest of the contributions using \eqref{eq:cyclicity} and \eqref{eq:simpleid}, we get
\beq
\begin{aligned}
\bar{d}_{12}^{\ell_{12}}\bar{d}_{34}^{\ell_{34}}\left[m\left(\frac{\chi-1}{\chi}\right)-4g^2 \left(L_R\left(\frac{x_{32}}{x_{31}}\right)+L_R\left(\frac{x_{21}}{x_{41}}\right)+L_R\left(\frac{x_{32}}{x_{42}}\right)+L_R\left(\frac{x_{43}}{x_{41}}\right)\right)\right]\period
\end{aligned}
\eeq
To further simplify this expression, we take the second equation of \eqref{eq:rogersid} and substitute $x$ with $x_{43}/x_{42}$ and $y$ with $x_{42}/x_{41}$. After this substitution, we get
\beq
\begin{aligned}
&0=L_{R}\left(\frac{x_{43}}{x_{42}}\right)+L_{R} \left(\frac{x_{42}}{x_{41}}\right)-L_R\left(\chi\right) -L_R\left(\frac{x_{43}}{x_{41}}\right)-L_R\left(\frac{x_{32}}{x_{31}}\right)\comma\\
&\iff L_{R}(\chi)-\frac{\pi^2}{3}=-\left[L_R\left(\frac{x_{32}}{x_{31}}\right)+L_R\left(\frac{x_{21}}{x_{41}}\right)+L_R\left(\frac{x_{32}}{x_{42}}\right)+L_R\left(\frac{x_{43}}{x_{41}}\right)\right]\period
\end{aligned}
\eeq
As a result we get the following one-loop correction:
\beq
{\tt boundary}|_{\ell_{13}=\ell_{24}=\ell_{41}=\ell_{23}=0}=\bar{d}_{12}^{\ell_{12}}\bar{d}_{34}^{\ell_{34}}\left[m\left(\frac{\chi-1}{\chi}\right)+4g^2 L_{R} (\chi)-\frac{4\pi^2 g^2}{3}\right]\period
\eeq

Let us next consider the case with $L_1+L_4>L_2+L_3$. In this case $\ell_{41}$ is nonzero, and the one-loop diagrams are given by (see also figure \ref{fig:sample}-$(b)$)
\beq
\begin{aligned}
&\ell_{12} {\sf S}_{12}\comma\quad (\ell_{12}-1)({\sf G}_{12|21}+{\sf Q}_{1221}) \comma\quad\ell_{34} {\sf S}_{34}\comma\quad (\ell_{34}-1)({\sf G}_{34|43}+{\sf Q}_{3443})\comma\\
&\ell_{41} {\sf S}_{41}\comma\quad (\ell_{41}-1)({\sf G}_{41|14}+{\sf Q}_{4114})\comma\quad {\sf G}_{12|34}+{\sf Q}_{1234}^{-}\comma\quad {\sf B}_{21;12}\comma\quad {\sf B}_{43;34}\comma\quad {\sf B}_{14;41}\comma\\
&{\sf B}_{12;23}\comma\quad {\sf B}_{34;23}\comma\quad {\sf B}_{41;23}\comma\quad {\sf G}_{21|14}+{\sf Q}_{2114}\comma\quad {\sf G}_{34|41}+{\sf Q}_{3441}\period
\end{aligned}
\eeq 
Summing these contributions, we get
\beq
\begin{aligned}
&\bar{d}_{12}^{\ell_{12}}\bar{d}_{34}^{\ell_{34}}\bar{d}_{41}^{\ell_{41}}\left[m\left(\frac{\chi-1}{\chi}\right)+4g^2 \left(L_{R}\left(\frac{x_{42}}{x_{41}}\right)-L_{R}\left(\frac{x_{43}}{x_{41}}\right)-L_{R}\left(\frac{x_{32}}{x_{42}}\right)-L_{R}\left(\frac{x_{32}}{x_{31}}\right)\right)\right]\period
\end{aligned}
\eeq
Using the identities for $L_{R}(x)$, we can rewrite it as
\beq
{\tt boundary}|_{\ell_{13}=\ell_{24}=\ell_{23}=0}=\bar{d}_{12}^{\ell_{12}}\bar{d}_{34}^{\ell_{34}}\bar{d}_{41}^{\ell_{41}}\left[m\left(\frac{\chi-1}{\chi}\right)+4g^2 L_{R} (\chi)-\frac{2\pi^2 g^2}{3}\right]\period
\eeq
Similarly, the case with $L_1+L_4<L_2+L_3$ ($\ell_{23}\neq 0$) can be computed, and the result reads
\beq
{\tt boundary}|_{\ell_{13}=\ell_{24}=\ell_{41}=0}=\bar{d}_{12}^{\ell_{12}}\bar{d}_{34}^{\ell_{34}}\bar{d}_{23}^{\ell_{23}}\left[m\left(\frac{\chi-1}{\chi}\right)+4g^2 L_{R} (\chi)-\frac{2\pi^2 g^2}{3}\right]\period
\eeq
\paragraph{Diagrams with diagonal contraction}
Next, we consider the diagrams with diagonal contractions. Here we focus on the case with $\ell_{13}\neq 0$ since the other case $\ell_{24}\neq 0$ can be obtained by the permutation of indices. 

As in the preceding analyses, we can list all the diagrams that contribute and sum them up with the help of identities such as \eqref{eq:cyclicity} and \eqref{eq:simpleid}. There is however much a simpler way to perform the computation in this case: Since all the diagrams locally look like the diagrams for the three-point functions (see figure \ref{fig:sample}-$(c)$), we can add and subtract the results for the three-point function, which we already computed. As a result we obtain
\beq
\begin{aligned}
&\left.{\tt boundary}\right|_{\ell_{23}=\ell_{41}=0,\,\ell_{13}\neq 0} =-\bar{d}_{12}^{\ell_{12}}\bar{d}_{34}^{\ell_{34}}\bar{d}_{13}^{\ell_{13}}\frac{2\pi^2g^2}{3}\comma\\
&\left.{\tt boundary}\right|_{\ell_{23}=0,\ell_{41}\neq 0,\,\ell_{13}\neq 0} =0\period
\end{aligned}
\eeq 
\subsection{Final results and comparison with localization}
Let us now finally write down the full answer for the one-loop four-point functions. Since we computed the corrections to the individual diagrams, what we need to do is to simply dress the tree-level answer \eqref{eq:treelevelfinal} by appropriate one-loop integrals. The result reads
\beq\label{eq:oneloopfinal}
\begin{aligned}
\left.\llangle\mathcal{O}_{L_1}\mathcal{O}_{L_2}\mathcal{O}_{L_3}\mathcal{O}_{L_4} \rrangle\right|_{O(g^2)}={\tt Left}+{\tt Bulk}+{\tt Right}\comma
\end{aligned}
\eeq
with
\beq\label{eq:decomposeoneloop1}
\begin{aligned}
&{\tt Bulk}=\\
&\delta_{\ell_{13}}\delta_{\ell_{24}}\left(m\left(\frac{\chi}{\chi-1}\right)+m\left(\frac{\chi-1}{\chi}\right)+\frac{2\pi^2 g^2}{3}\right){\sf D}_{\rm left}\sum_{n=1}^{n_{\rm max}-1}\left[\frac{\chi^{2}}{(1-\chi)^2}\frac{(1-\alpha)(1-\bar{\alpha})}{\alpha\bar{\alpha}}\right]^{n}\comma\\
&{\tt Left}={\sf D}_{\rm left}\left[-\frac{2\pi^2 g^2\delta_{\ell^{\rm min}_{23}}\delta_{\ell^{\rm min}_{41}}}{3}+\delta_{\ell_{13}}\delta_{\ell_{24}}\left(m\left(\frac{\chi-1}{\chi}\right)+4g^2 L_{R} (\chi)-\frac{2\pi^2 g^2}{3}\right)\right]\comma\\
&{\tt Right}={\sf D}_{\rm right}\left[-\frac{2\pi^2 g^2\delta_{\ell^{\rm min}_{12}}\delta_{\ell^{\rm min}_{34}}}{3}+\delta_{\ell_{13}}\delta_{\ell{24}}\left(m\left(\frac{\chi}{\chi-1}\right)+4g^2 L_{R} (1-\chi)-\frac{2\pi^2 g^2}{3}\right)\right]\period
\end{aligned}
\eeq
Here $\delta_{\ell_{ij}}$ means the Kronecker delta $\delta_{\ell_{ij},0}$.

We can also compute the normalized four-point functions by dividing the answer by $\sqrt{n_{L_1}n_{L_2}n_{L_3}n_{L_4}}$. The answer reads
\beq\label{eq:oneloopfinal2}
\left.\frac{\llangle\mathcal{O}_{L_1}\mathcal{O}_{L_2}\mathcal{O}_{L_3}\mathcal{O}_{L_4} \rrangle}{\sqrt{n_{L_1}n_{L_2}n_{L_3}n_{L_4}}}\right|_{O(g^2)}={\tt Left}^{\prime}+{\tt Bulk}^{\prime}+{\tt Right}^{\prime}\comma
\eeq
with
\beq\label{eq:decomposeoneloop2}
\begin{aligned}
&{\tt Bulk}^{\prime}=\\
&\delta_{\ell_{13}}\delta_{\ell_{24}}\left(m\left(\frac{\chi}{\chi-1}\right)+m\left(\frac{\chi-1}{\chi}\right)+2\pi^2 g^2\right){\sf D}^{\prime}_{\rm left}\sum_{n=1}^{n_{\rm max}-1}\left[\frac{\chi^{2}}{(1-\chi)^2}\frac{(1-\alpha)(1-\bar{\alpha})}{\alpha\bar{\alpha}}\right]^{n}\comma\\
&{\tt Left}^{\prime}={\sf D}^{\prime}_{\rm left}\left[\frac{4\pi^2g^2}{3}-\frac{2\pi^2 g^2\delta_{\ell^{\rm min}_{23}}\delta_{\ell^{\rm min}_{41}}}{3}+\delta_{13}\delta_{24}\left(m\left(\frac{\chi-1}{\chi}\right)+4g^2 L_{R} (\chi)-\frac{2\pi^2 g^2}{3}\right)\right]\comma\\
&{\tt Right}^{\prime}={\sf D}^{\prime}_{\rm right}\left[\frac{4\pi^2g^2}{3}-\frac{2\pi^2 g^2\delta_{\ell^{\rm min}_{12}}\delta_{\ell^{\rm min}_{34}}}{3}+\delta_{13}\delta_{24}\left(m\left(\frac{\chi}{\chi-1}\right)+4g^2 L_{R} (1-\chi)-\frac{2\pi^2 g^2}{3}\right)\right]\period
\end{aligned}
\eeq
Here ${\sf D}^{\prime}_{\rm left,right}$ are defined by
\beq
\begin{aligned}
&{\sf D}^{\prime}_{\rm left}=d_{12}^{\ell^{\rm max}_{12}}d_{34}^{\ell^{\rm max}_{34}}d_{41}^{\ell^{\rm min}_{41}}d_{23}^{\ell^{\rm min}_{23}}d_{13}^{\ell_{13}}d_{24}^{\ell_{24}}\comma\qquad 
&{\sf D}^{\prime}_{\rm right}=d_{12}^{\ell^{\rm min}_{12}}d_{34}^{\ell^{\rm min}_{34}}d_{41}^{\ell^{\rm max}_{41}}d_{23}^{\ell^{\rm max}_{23}}d_{13}^{\ell_{13}}d_{24}^{\ell_{24}}\period
\end{aligned}
\eeq

Let us now compare the results we obtained with the results of the localization computation \cite{Giombi:2018qox}. For this purpose, we need to set
\beq
d_{ij}=-\frac{1}{2}\comma\qquad \chi=\alpha=\bar{\alpha}\period
\eeq
We then get
\beq
\begin{aligned}
\left.{\tt Bulk}\right|_{\rm localization}&=\left(-g^2\right)^{\frac{L_1+L_2+L_3+L_4}{2}}\frac{2\pi^2g^2\delta_{\ell_{13}}\delta_{\ell_{24}}}{3}(n_{\rm max}-1)\comma\\
\left.{\tt Left}+{\tt Right}\right|_{\rm localization}&=-\left(-g^2\right)^{\frac{L_1+L_2+L_3+L_4}{2}}\frac{2\pi^2g^2(\delta_{\ell_{13}}\delta_{\ell_{24}}+\delta_{\ell^{\rm min}_{12}}\delta_{\ell^{\rm min}_{34}}+\delta_{\ell^{\rm min}_{23}}\delta_{\ell^{\rm min}_{41}})}{3}\period
\end{aligned}
\eeq
We compared the sum of these two results with the results obtained by the localization. In all the cases that we tested, the results agreed perfectly.
\section{Hexagonalization\label{sec:hexagonalization}}
Having computed the one-loop corrections, we now discuss how the results could be reproduced using integrability. 
\subsection{Proposal}
For the correlation functions of single-trace operators, the integrability-based approach, called {\it hexagonalization}, was proposed in \cite{Fleury:2016ykk}. The key idea of the hexagonalization is to cut the worldsheet into hexagonal patches and then glue them back together multiplying appropriate weight factors.
  \begin{figure}[t]
\centering
\includegraphics[clip,height=6cm]{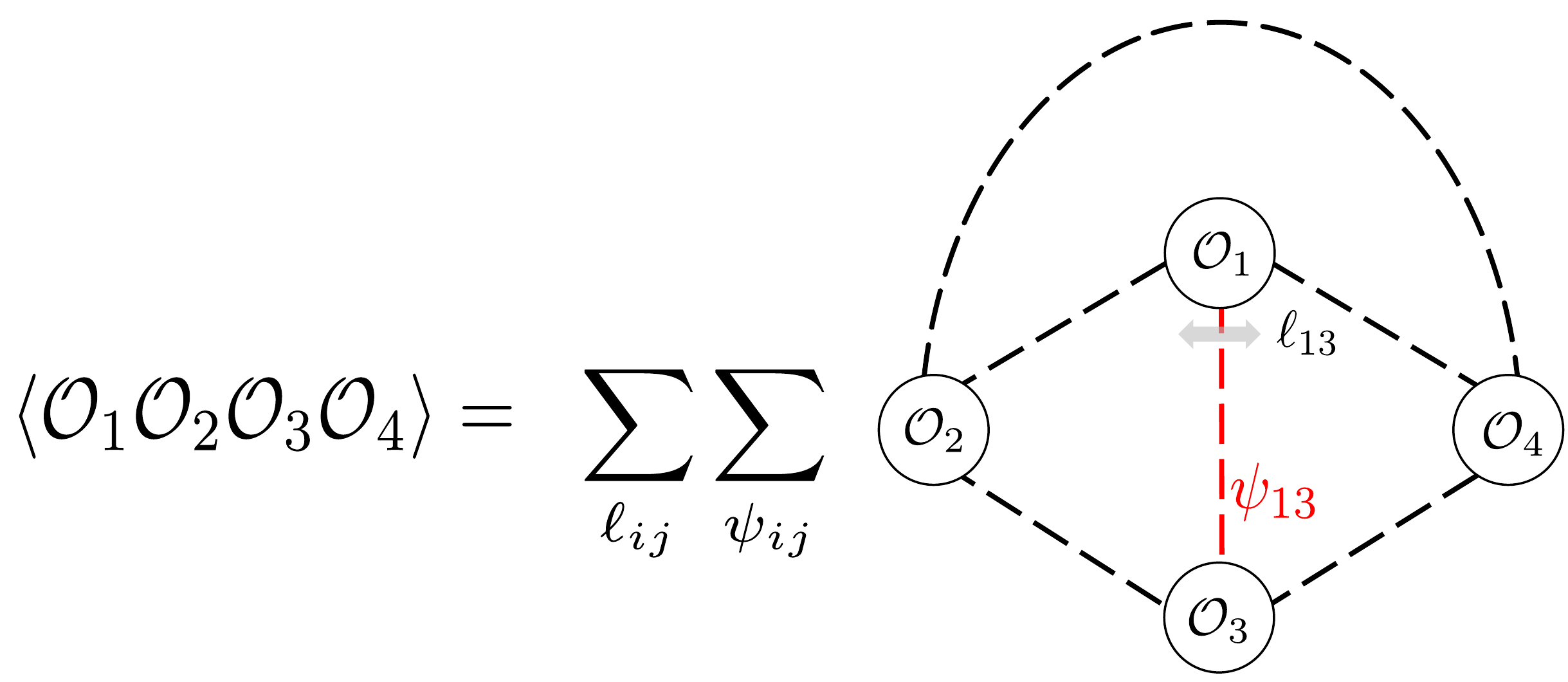}
\caption{Hexagonalization of the four-point functions. We draw all tree-level graphs, decorate them by inserting nontrivial states on the dashed edges and compute the contribution from each hexagon.}
\label{fig:closedhex}
\end{figure}

The approach consists of three steps: We first draw all possible tree-level graphs, each of which is distinguished by the numbers of contractions between operators. Next, we tessellate each graph into hexagons. The last step, which is the most important, is to glue the hexagons back together by summing over all possible states that appear on the glued edges. This last step is reminiscent of the sewing construction in 2d CFT, the difference being that the states that appear in our case are labelled by the numbers and the momenta of excitations (called magnons). The dependence on the cross ratio is incorporated as the \textcolor[rgb]{0,0,0.7}{\it weight factor} for each state and the bridge-lengths appear as the \textcolor[rgb]{0,0.4,0}{\it propagation factor} for individual magnons. For the planar four-point function, these procedures lead to the following expression\fn{Note that here we included the measure factors into the hexagon form factor $\mathcal{H}$. More precisely the relation between the definition in \cite{Fleury:2016ykk} and the definition here is given by
\beq
\left.\mathcal{H}_{\psi_1,\psi_2,\psi_3}\right|_{\rm here}=\sqrt{\mu_{\psi_1}\mu_{\psi_2}\mu_{\psi_3}}\left.\mathcal{H}_{\psi_1,\psi_2,\psi_3}\right|_{\rm there}\period
\eeq
For explicit expressions of various factors, see \cite{Fleury:2016ykk}. } :
\beq\label{eq:general}
\begin{aligned}
\langle \mathcal{O}_{L_1}\mathcal{O}_{L_2}\mathcal{O}_{L_3}\mathcal{O}_{L_4}\rangle=\sum_{{\rm graphs}}\underbrace{\prod_{(i,j)} (d_{ij})^{\ell_{ij}}}_{\text{tree level}}\left[ \sum_{\{\psi_{ij}\}}\prod_{(i,j)}\,\,\textcolor[rgb]{0,0.4,0}{\underbrace{e^{-\tilde{E}_{\psi_{ij}} \ell_{ij}}}_{\text{propagation}}}\,\,\textcolor[rgb]{0,0,0.7}{\underbrace{\mathcal{W}_{\psi_{ij}}}_{\text{weight}}} \,\,\prod_{(i,j,k)} \textcolor[rgb]{1,0,0}{\underbrace{\mathcal{H}_{\psi_{ij},\psi_{jk},\psi_{ki}}}_{\text{hexagon}}}\right]\period
 \end{aligned}
\eeq
See also figure \ref{fig:closedhex} for a pictorial explanation.

 \begin{figure}[t]
\centering
\includegraphics[clip,height=5cm]{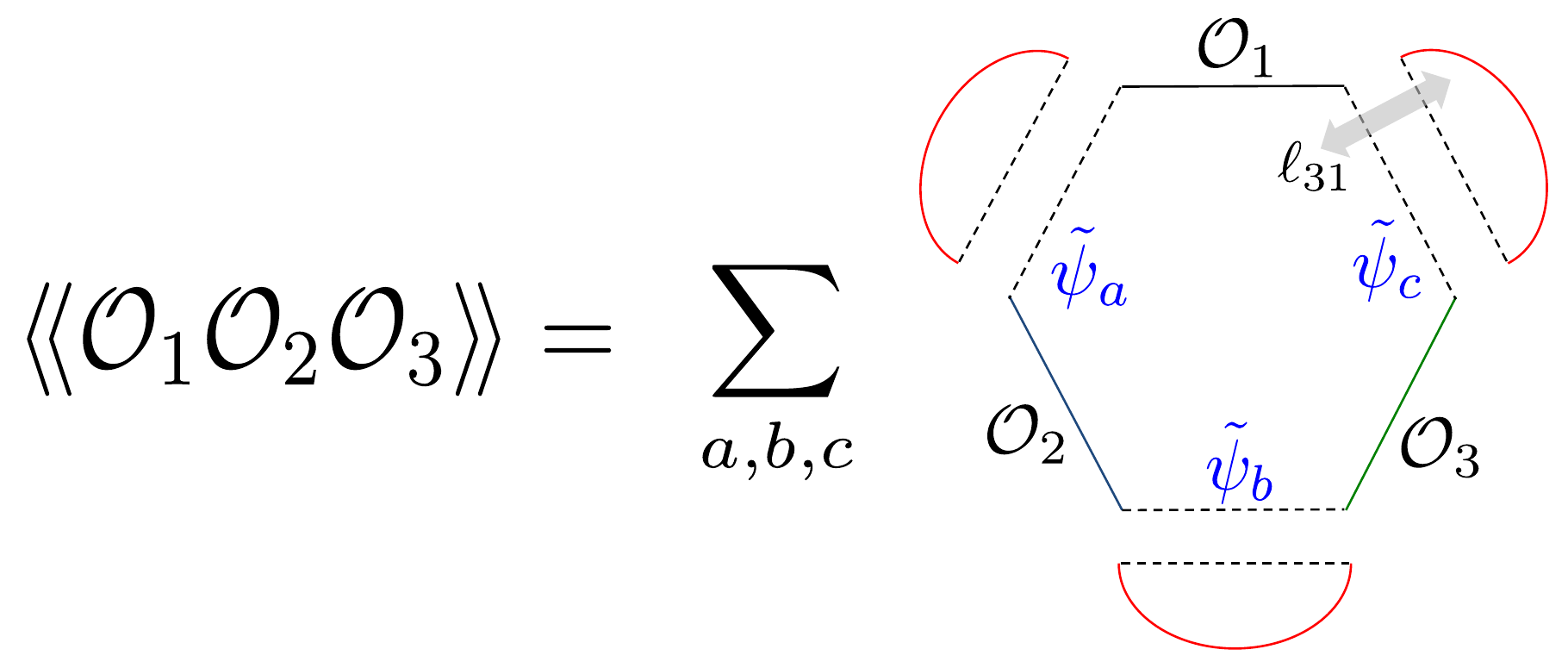}
\caption{Hexagonalization of the three-point function on the Wilson loop. The red curves denote the boundary states. We glue the boundary states to the hexagon by summing over all the intermediate states that appear in the expansion of the boundary state \eqref{eq:expansionofBstate}.}
\label{fig:openhex1}
\end{figure}

The goal of this section is to propose a generalization of this approach to the correlators on the Wilson loop. The planar diagrams for the correlators on the Wilson loop have the topology of a disk, and can also be tessellated into hexagons as shown in figure \ref{fig:openhex1}. The main difference, as compared to the correlators of single-trace operators, is that some of the edges of the hexagons are attached to the segments of the Wilson loop. In the integrability description, it is known \cite{Drukker:2012de,Correa:2012hh} that the Wilson loop is described by a boundary state, whose structure can be expressed schematically as
\beq
| \mathcal{B} \rangle \sim \exp \left[\int dp \, K^{ab} (p)A_a^{\dagger} (-p)A_b^{\dagger} (p) \right]|\Omega \rangle\comma
\eeq
where $A_{a}^{\dagger}$ is the creation operator of a magnon with index $a$ and $K^{ab}(p)$ is the analytically continued reflection matrix (see \cite{Drukker:2012de,Correa:2012hh} for explicit expressions). Expanding the exponential, the boundary state will be given by a sum of multi-particle states $|\tilde{\psi}_k\rangle$ with definite momentum and energy:
\beq\label{eq:expansionofBstate}
|\mathcal{B}\rangle =\sum_{k}b_{\tilde{\psi}_k}|\tilde{\psi}_k\rangle\comma
\eeq
where $b_{\tilde{\psi}_k}$ is a numerical coefficient.

\begin{figure}[t]
\centering
\includegraphics[clip,height=5cm]{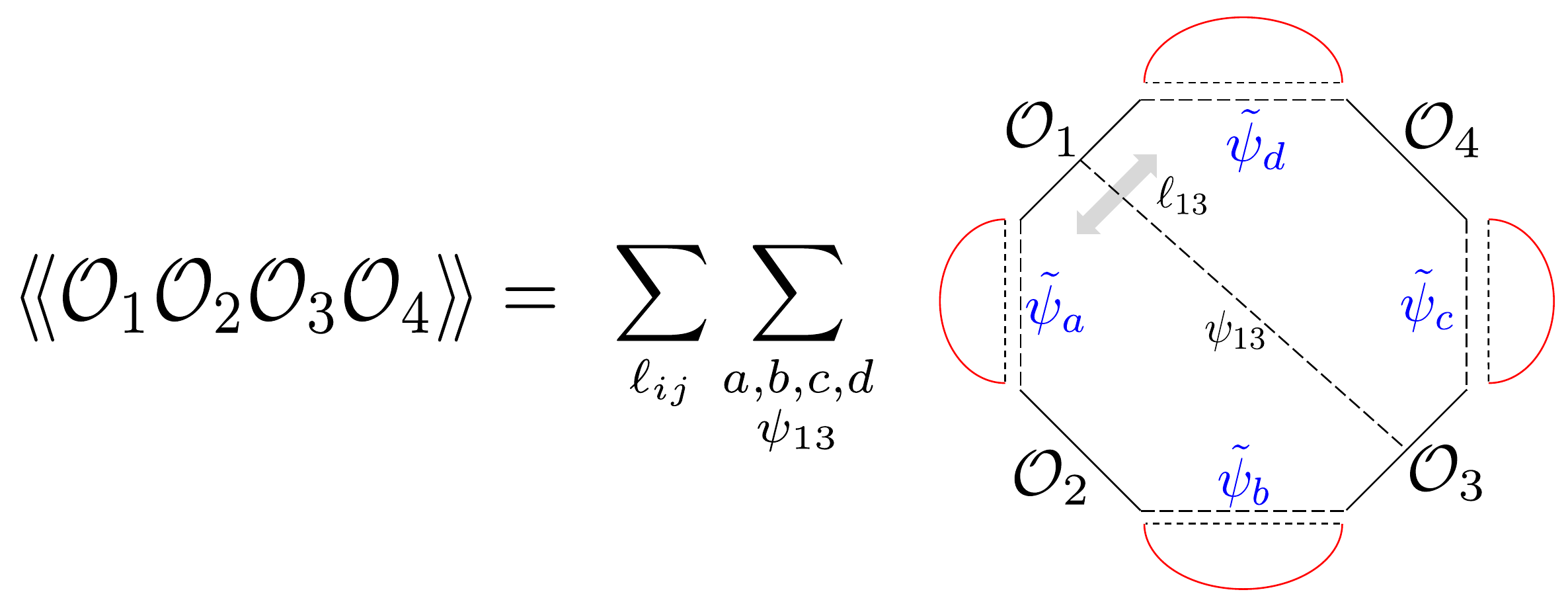}
\caption{Hexagonalization of the four-point function on the Wilson loop. In this case, we need to glue together two hexagons and four boundary states. This results in the sum over the boundary states $\tilde{\psi}$'s and the sum over the states $\psi_{13}$ that stitch the two hexagons.}
\label{fig:openhex2}
\end{figure}

Given such a structure, it is natural to conjecture that the edges of the hexagon attached to the Wilson loop are contracted with the boundary state. For instance, the conjecture for the three-point function of protected operators on the Wilson loop reads\fn{Note that, since the three-point function does not have any cross ratios, the weight factor is just $1$ for the three-point function.} (see also figure \ref{fig:openhex1})
\beq\label{eq:3pthex}
\frac{\llangle \mathcal{O}_{L_1}\mathcal{O}_{L_2}\mathcal{O}_{L_3}\rrangle}{\sqrt{n_{L_1}n_{L_2}n_{L_3}}} \overset{?}{=} d_{12}^{\ell_{12}}d_{23}^{\ell_{23}}d_{31}^{\ell_{31}}\sum_{a,b,c}e^{-(\tilde{E}_{\textcolor[rgb]{0,0,1}{\tilde{\psi}_{a}}}\ell_{12}+\tilde{E}_{\textcolor[rgb]{0,0,1}{\tilde{\psi}_{b}}}\ell_{23}+\tilde{E}_{\textcolor[rgb]{0,0,1}{\tilde{\psi}_{c}}}\ell_{31})}b_{\textcolor[rgb]{0,0,1}{\tilde{\psi}_a}}b_{\textcolor[rgb]{0,0,1}{\tilde{\psi}_b}}b_{\textcolor[rgb]{0,0,1}{\tilde{\psi}_c}}\mathcal{H}_{\textcolor[rgb]{0,0,1}{\tilde{\psi}_{a}},\textcolor[rgb]{0,0,1}{\tilde{\psi}_{b}},\textcolor[rgb]{0,0,1}{\tilde{\psi}_{c}}}\period
\eeq 
Note that $\textcolor[rgb]{0,0,1}{\tilde{\psi}}$'s here are the states obtained by the expansion of the boundary state.
Similarly the conjecture for the four-point function is given by\fn{Here we assumed that $\ell_{24}=0$. The generalization for $\ell_{24}\neq 0$ is straightforward.} (see also figure \ref{fig:openhex2})
\beq\label{eq:4pthex}
\begin{aligned}
&\frac{\llangle\mathcal{O}_{L_1}\mathcal{O}_{L_2}\mathcal{O}_{L_3}\mathcal{O}_{L_4} \rrangle}{\sqrt{n_{L_1}n_{L_2}n_{L_3}n_{L_4}}}\overset{?}{=}\sum_{\ell_{ij}}\prod_{(i,j)}(d_{ij})^{\ell_{ij}}\\
&\sum_{\substack{a,b,c,d\\ \psi_{13}}}e^{-(\tilde{E}_{\textcolor[rgb]{0,0,1}{\tilde{\psi}_{a}}}\ell_{12}+\tilde{E}_{\textcolor[rgb]{0,0,1}{\tilde{\psi}_{b}}}\ell_{23}+\tilde{E}_{\textcolor[rgb]{0,0,1}{\tilde{\psi}_{c}}}\ell_{34}+\tilde{E}_{\textcolor[rgb]{0,0,1}{\tilde{\psi}_{d}}}\ell_{41})}e^{-\tilde{E}_{\psi_{13}}\ell_{13}}\mathcal{W}_{\psi_{13}}b_{\textcolor[rgb]{0,0,1}{\tilde{\psi}_{a}}}b_{\textcolor[rgb]{0,0,1}{\tilde{\psi}_{b}}}b_{\textcolor[rgb]{0,0,1}{\tilde{\psi}_{c}}}b_{\textcolor[rgb]{0,0,1}{\tilde{\psi}_{d}}}\mathcal{H}_{\textcolor[rgb]{0,0,1}{\tilde{\psi}_{a}},\textcolor[rgb]{0,0,1}{\tilde{\psi}_b}, \psi_{13}}\mathcal{H}_{\psi_{13},\textcolor[rgb]{0,0,1}{\tilde{\psi}_{c}},\textcolor[rgb]{0,0,1}{\tilde{\psi}_d}}\period
\end{aligned}
\eeq

In this paper, we will not perform the explicit integrability computation. Instead we will compare the perturbative results obtained in the preceding sections with the conjectures put forward above. As we see below, the comparison leads to a modification of the proposal.
\subsection{A puzzle and a resolution}
The proposal above allows us to compute the correction to each tree-level diagram. On the other hand, the one-loop results computed in the previous section are also given as the corrections to the individual diagrams. Therefore one can compare the one-loop results with the integrability computation at the level of individual diagrams.

\paragraph{Puzzle} Such comparison, however, leads to some puzzle. To see this, we need to recall how each factor in \eqref{eq:3pthex} and \eqref{eq:4pthex} scales at weak coupling. As discussed in \cite{Basso:2015zoa}, the propagation factor scales as
\beq\label{eq:suppression}
e^{-\tilde{E}\ell}=O(g^{2\ell})\comma
\eeq
while the $n$-particle state contribution to the boundary state scales as (see \cite{Drukker:2012de,Correa:2012hh})
\beq
b_{\psi_{\text{n-particle}}}=O(g^{n})\period
\eeq
In addition, the hexagon form factor for the $n$-particle state (and two vacuum states $\emptyset$) scales as
\beq
\mathcal{H}_{\psi_{\text{n-particle}},\emptyset,\emptyset}=O(g^{n})\period
\eeq
Here, we only considered the hexagon form factor with one excited state, but the analysis can be readily generalized to the case with multiple excited states. As a result, one can show that the only configurations that contribute at one loop are one-particle states on zero-length edges ($\ell_{ij}$), or ``multi-particle string'' configurations discussed in \cite{Bargheer:2018jvq}. 

This in particular means that, for three-point functions with $\ell_{ij}\neq 0$, there is no configuration that contributes at one loop. However, this is clearly in contradiction with the results we obtained \eqref{eq:3ptfinal}, which are nonzero $(\pi^2g^2)$ even when all $\ell_{ij}$'s are nonzero. A similar contradiction can be seen also in the four-point function. For instance, consider the bulk diagram without diagonal contractions, namely $\ell_{13}=\ell_{24}=0$. The relevant magnon configuration for this diagram is the one-particle state living on the (zero-length) diagonal edge. This is precisely the configuration analyzed in \cite{Fleury:2016ykk}, and the result is simply given by
\beq
\left.{\tt bulk}\right|_{\ell_{13}=\ell_{24}=0}=\bar{d}_{12}^{\ell_{12}}\bar{d}_{23}^{\ell_{23}}\bar{d}_{34}^{\ell_{34}}\bar{d}_{41}^{\ell_{41}}\left[m\left(\frac{\chi}{\chi-1}\right)+m\left(\frac{\chi-1}{\chi}\right)\right]\qquad (\text{integrability})\period
\eeq
Comparing this with the perturbative results, \eqref{eq:decomposeoneloop1} and \eqref{eq:decomposeoneloop2}, we see discrepancies by a constant shift ($\frac{2\pi^2g^2}{3}$ or $2\pi^2 g^2$).

\paragraph{Resolution} Fortunately, there is a simple resolution to this puzzle. So far we have been comparing the integrability prediction with the correlators on the straight-line Wilson loop since they have direct connections with the defect CFT data. However, it turns out that the integrability computation is more naturally related to the {\it normalized} correlator on the circular Wilson loop defined by
\beq\label{eq:correspondence}
\frac{\langle \mathcal{W} [\mathcal{O}_1\cdots \mathcal{O}_m]\rangle_{\rm circle}}{\prod_{k=1}^{m}\sqrt{\langle \mathcal{W}[\mathcal{O}_k\mathcal{O}_k]\rangle_{\rm circle}^{\prime}}}=\left(\langle \mathcal{W} \rangle_{\rm circle}\right)^{\frac{2-m}{2}} \frac{\llangle\mathcal{O}_1\cdots \mathcal{O}_m \rrangle}{\sqrt{\prod_{k=1}^{m}n_k}}\period
\eeq
Using the perturbative answer for the expectation value of the circular Wilson loop \cite{Drukker:2000rr},
\beq
\langle\mathcal{W}\rangle_{\rm circle}=1+2\pi^{2}g^2 +O(g^{4})\comma
\eeq
one can show that the extra factor $\left(\langle \mathcal{W} \rangle_{\rm circle}\right)^{\frac{2-m}{2}}$ kills precisely the unnecessary constant shifts. 

\paragraph{Large Charge Asymptotics}We do not have a clear physical explanation as to why the integrability computation seems to give the correlators on the circular Wilson loop rather than the straightline Wilson loop. One interesting outcome of this conjecture is that the large-charge asymptotics of the structure constant is given by the expectation value of the Wilson loop. To see this, consider the three-point functions of long operators; more precisely the three-point function with $\ell_{ij}\gg 1$. Owing to the suppression coming from the propagation factor, the contributions from nontrivial magnon configurations are all suppressed, which means that the integrability answer is just $1$. Using the correspondence \eqref{eq:correspondence}, this translates to the following asymptotics of the structure constant:
\beq
C_{L_1,L_2,L_3}\quad \overset{\ell_{ij}\gg 1}{\sim}\quad \sqrt{\langle \mathcal{W}\rangle_{\rm circle}}=\left(\frac{2 I_1 (\sqrt{\lambda})}{\sqrt{\lambda}}\right)^{\frac{1}{2}}\period 
\eeq
In principle, this prediction can be verified by the localization computation \cite{Giombi:2018qox}. It would also be interesting if one could derive this asymptotics from the defect CFT point of view\fn{See e.g.~\cite{Hellerman:2015nra} for the analysis of the CFT data in the large charge limit.}.
\paragraph{Comparison with Localization}
To provide further evidence for the conjecture, let us analyze the results computed by the localization in \cite{Giombi:2018qox}. We refer the original paper for the details of the computation and simply show the results for a few short operators:
\beq
\begin{aligned}
&\tilde{C}_{2,1,1}=1-\frac{2\pi^2g^2}{3}+O(g^{4})\comma\qquad
&&\tilde{C}_{2,2,2}=1-\frac{14\pi^4g^{4}}{15}+O(g^{6})\comma\\&\tilde{C}_{3,2,1}=1-\frac{2\pi^2g^2}{3}+O(g^{4})\comma\qquad
&&\tilde{C}_{4,4,4}=1-\frac{124\pi^{6}g^{6}}{315}+O(g^{8})\comma
\end{aligned}
\eeq
where $\tilde{C}_{L_1,L_2,L_3}$ is the normalized structure constant on the circular Wilson loop,
\beq
\tilde{C}_{L_1,L_2,L_3}\equiv \left(\langle \mathcal{W}\rangle_{\rm circle}\right)^{-\frac{1}{2}}C_{L_1,L_2,L_3}\period
\eeq
One can readily verify that the leading correction shows up at order\fn{Note that, for three-point functions, the number of contractions between operators is determined completely by the lengths of the operators, $\ell_{ij}=(L_i+L_j-L_k)/2$.} 
\beq
O\left(g^{2 {\rm min}\left[\ell_{12},\ell_{23},\ell_{31}\right]}\right)\comma
\eeq
as is consistent with the suppression due to the propagation factor \eqref{eq:suppression}.
\subsection{Prediction for multi-particle contributions}
Let us now perform a more detailed comparison between the integrability computation and the one-loop results for the three- and four-point functions, and make predictions for multi-particle contributions in the hexagonalization approach.

First, by comparing the results for the three-point function, one can immediately conclude that the contribution from the one-particle state living on a boundary edge, namely an edge of a hexagon attached to the boundary state, is given by
\beq\label{eq:readingoff1}
\text{\tt (1-particle)}=-\frac{2\pi^2g^2}{3}\period
\eeq

\begin{figure}[t]
\centering
\includegraphics[clip,height=6cm]{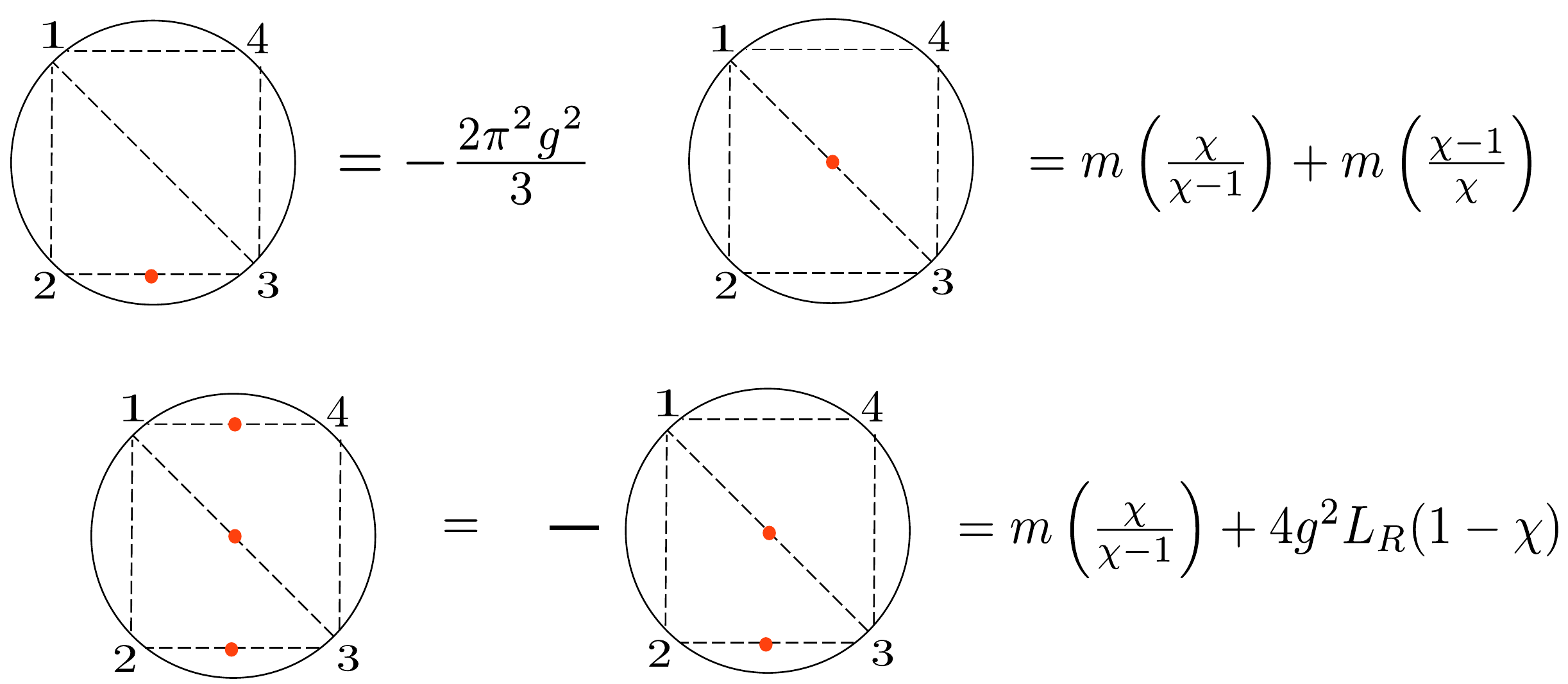}
\caption{Various magnon configurations and their contributions. The red dots denote the magnons. The first configuration is $(\text{\tt 1-particle})$ while the second is $(\text{\tt bulk 1-particle})$. The third and the fourth configurations are $(\text{\tt 3-particle})$ and $(\text{\tt 2-particle})$ respectively.}
\label{fig:magnons}
\end{figure}
Let us next analyze the four-point functions. At one loop, there are several relevant configurations that contribute to the four-point functions; the one-particle states, the two-particle state, and the three-particle state. Among those contributions, we already know the contribution from the one-particle state is given either by \eqref{eq:readingoff1} or by
\beq
\text{\tt (bulk 1-particle)}=m\left(\frac{\chi}{\chi-1}\right)+m \left(\frac{\chi-1}{\chi}\right)\comma
\eeq
depending on whether the magnon lives on the boundary edge or on the diagonal edge. By subtracting these contributions from the one-loop answer, one can determine the contributions from the rest of the configurations. For instance, the diagram with $\ell_{13}=\ell_{24}=\ell_{23}=0$  and nonzero $\ell_{12}$, $\ell_{34}$ and $\ell_{41}$ receive the contributions from three different configurations,
\beq
\text{\tt (1-particle)}\comma\quad \text{\tt (2-particle)} \comma\quad 
\text{\tt (bulk 1-particle)}\comma
\eeq 
and their sum is given by\fn{See \eqref{eq:decomposeoneloop2}.}
\beq
-\frac{2\pi^2g^2}{3}+m\left(\frac{\chi-1}{\chi}\right)+4g^2L_{R}(\chi)-\frac{2\pi^2g^2}{3}\period
\eeq
We therefore conclude that the contribution from the two-particle state is given by
\beq
\begin{aligned}
\text{\tt (2-particle)}=&-m\left(\frac{\chi}{\chi-1}\right)+4g^2L_{R}(\chi)-\frac{2\pi^2g^2}{3}\\
=&-m\left(\frac{\chi}{\chi-1}\right)-4g^2 L_{R} (1-\chi)\period
\end{aligned}
\eeq
Similarly, one can determine the contribution from the three-particle state as
\beq
\begin{aligned}
\text{\tt (3-particle)}=&m\left(\frac{\chi}{\chi-1}\right)-4g^2L_{R}(\chi)+\frac{2\pi^2g^2}{3}\\
=&m\left(\frac{\chi}{\chi-1}\right)+4g^2 L_{R} (1-\chi)\period
\end{aligned}
\eeq 
See figure \ref{fig:magnons} for a summary of the results.

Obviously, an important question is to reproduce these results by performing an explicit integrability computation. Since both the boundary state and the hexagon form factors are known, this is in principle a possible task. It can however be technically complicated and we will leave it for future investigations.
\section{Non-BPS Structure Constants\label{sec:nonBPS3pt}}
In the previous section, we have seen that the hexagonalization computes the correlators of the BPS operators on the {\it circular} Wilson loop. It is then natural to conjecture that the same holds also for the correlators of the non-BPS operators.

In general, the three-point functions on the straightline Wilson loop in the planar limit are given by
\beq
\llangle \mathcal{O}_1 (x_1)\mathcal{O}_2 (x_2)\mathcal{O}_3 (x_3)\rrangle=\frac{1}{\sqrt{N}}\frac{C_{123}}{|x_{12}|^{\Delta_1+\Delta_2-\Delta_3}|x_{23}|^{\Delta_2+\Delta_3-\Delta_1}|x_{31}|^{\Delta_3+\Delta_1-\Delta_2}}\period
\eeq
Here we assumed that the two-point function of each operator is canonically normalized. In order to convert this into the correlator on the circular Wilson loop, we need to multiply $(\langle \mathcal{W}\rangle_{\rm circle})^{-\frac{1}{2}}$  as in \eqref{eq:correspondence} and define
\beq
\tilde{C}_{123}\equiv (\langle \mathcal{W}\rangle_{\rm circle})^{-\frac{1}{2}}\,\,C_{123}\period
\eeq
 Then the conjecture is that it is $\tilde{C}_{123}$ (rather than $C_{123}$) that can be computed by the integrability method.
\paragraph{Asymptotic part}
In general, the hexagon approach to the non-BPS three-point function consists of two parts, the asymptotic part and the finite-size corrections.

The expression for the asymptotic part of the structure constant with a single non-BPS operator in the rank $1$ sector was given in \cite{Kim:2017phs}. The only difference from the current analysis is that, in \cite{Kim:2017phs}, it was conjectured that the integrability method computes the standard structure constant $C_{123}$. However, as we saw above, in view of the comparison of the one-loop results, it is more natural to conjecture that the integrability method computes $\tilde{C}_{123}$.

Having made this remark, let us briefly review the proposal in \cite{Kim:2017phs} (with the modification mentioned above). The proposal is a natural generalization of the one made in \cite{Basso:2015zoa}, and reads\fn{Note that the paper \cite{Kim:2017phs} uses a convention for the S-matrix which is not standard in the $\mathcal{N}=4$ SYM literature. Here we rewrote it using standard conventions:
\beq
\left.S(u_i,u_j)\right|_{\rm here}=\left. S(u_j,u_i)\right|_{\text{in \cite{Kim:2017phs}}}\period
\eeq}
\beq
(\tilde{C}_{123})^2=\frac{e^{i p_{\rm tot}(\ell_{12}-\ell_{31})}}{\det (\del_{u_i}\phi_j) \prod_{i<j}S(u_i,u_j)}(\mathcal{A})^2\period
\eeq
Here $\phi_j$ is the phase factor\fn{The Bethe equation is given by $e^{i\phi_j}=1$.}
\beq
e^{i\phi_j} \equiv e^{2 i p_j L} R_{R}(u_j)R_{L}(\bar{u}_j)\prod_{k\neq j}S(u_j,u_k)S(u_k,\bar{u}_j)\comma
\eeq
where $S(u,v)$ is the S-matrix while $R_L$ and $R_R$ are the reflection matrices for the left- and the right-boundaries of the spin chain. The barred rapidities $\bar{u}$ denote the parity-flipped rapidities which can be characterized by the values of the Zhukovski variables\fn{The Zhukovski variables are defined by
\beq
x(u)+\frac{1}{x (u)}=\frac{u}{g}\comma\qquad x^{\pm}(u)=x\left(u\pm\tfrac{i}{2}\right)\period
\eeq
}
\beq
x^{+}(\bar{u})=-x^{-}(u)\comma\qquad x^{-}(\bar{u})=-x^{+}(u)\comma
\eeq
and $p_{\rm tot}$ is the total momentum
\beq
p_{\rm tot}\equiv \sum_j p_j\period
\eeq
The main part of the conjecture $\mathcal{A}$ is a sum over partitions defined by
\beq\label{eq:sumoverpartitions}
\mathcal{A}=\sum_{\alpha_{+}\cup \alpha_{-}=\{1,\dots, M\}}\left[\prod_{j\in \alpha_{-}}\left(-e^{2i p_{j}\ell_{31}}R_{R}(u_j)\right)\prod_{k>j}S(u_j,u_k)S(u_k,\bar{u}_j)\right]\prod_{s<t}h (\hat{u}_s ,\hat{u}_t)\period
\eeq
Here $h(u,v)$ are the hexagon form factors and $\hat{u}_s$'s are given by
\beq
\hat{u}_i =\begin{cases}u_i\qquad &i\in \alpha_{+}\\\bar{u}_i\qquad &i\in \alpha_{-}\end{cases}\period
\eeq
\begin{figure}[t]
\centering
\includegraphics[clip,height=5cm]{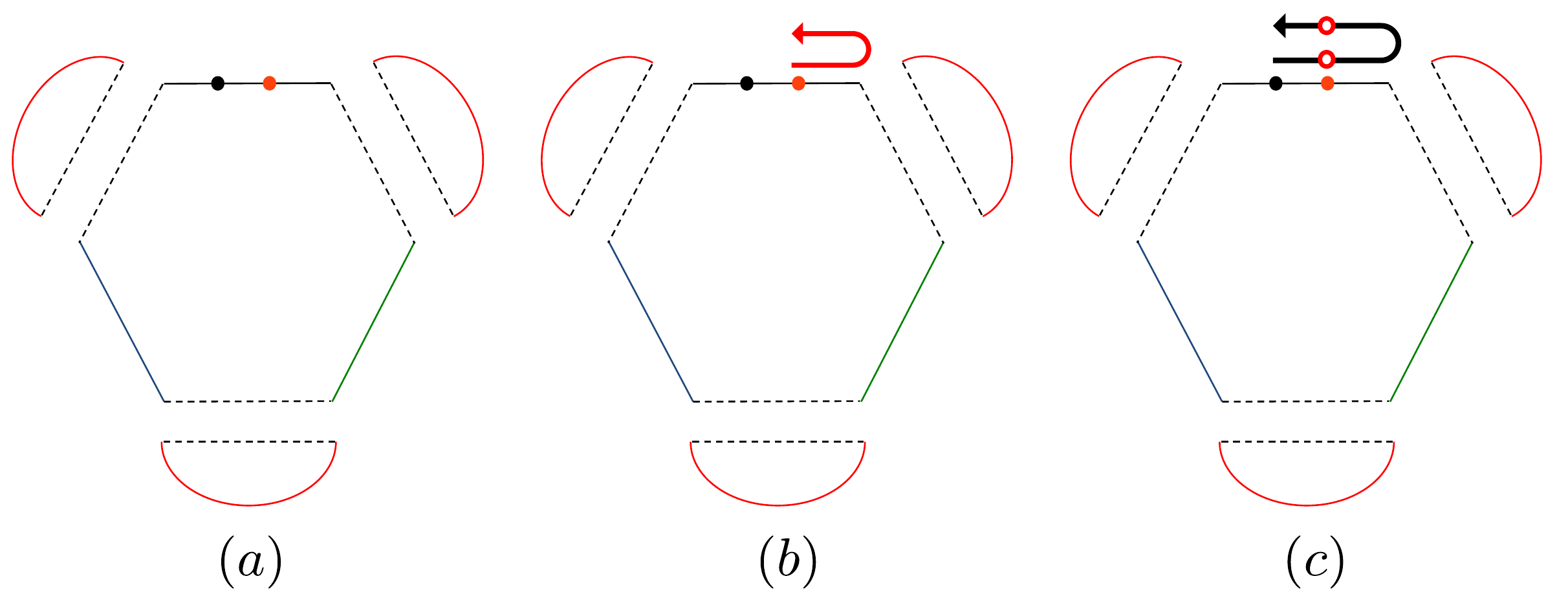}
\caption{Physical interpretation of the sum over partitions. The black dot denotes a magnon with $u_1$ and the red dot denotes a magnon with $u_2$. $(a)$ The first term in the sum \eqref{eq:sumoverpartitionsreal}, which is just a hexagon form factor with $u_1$ and $u_2$. $(b)$ The second term in the sum \eqref{eq:sumoverpartitionsreal}. To flip the momentum, the second magnon has to travel to the right boundary and get reflected. This leads to the propagation factor shown in \eqref{eq:sumoverpartitionsreal}. $(c)$ The third term in the sum \eqref{eq:sumoverpartitionsreal}. In this case, the first magnon needs to travel to the right edge and get reflected. In the process, it passes through the second magnon twice, first from the left and second from the right. This leads to additional S-matrix factors.}
\label{fig:partition}
\end{figure}

Although the expression \eqref{eq:sumoverpartitions} might look complicated at first sight, it has a natural physical interpretation. To see this, let us consider the expression for the two-magnon state:
\beq\label{eq:sumoverpartitionsreal}
\begin{aligned}
\mathcal{A}=&h(u_1,u_2)-e^{2i p_2 \ell_{31}}R_{R}(u_2)h(u_1,\bar{u}_2)-e^{2i p_1 \ell_{31}}S(u_1,u_2)S(u_2,\bar{u}_1)h(\bar{u}_1,u_2)\\
&+e^{2i (p_1+p_2)\ell_{31}} R_{R}(u_1)R_{R}(u_2)S(u_1,u_2)S(u_2,\bar{u}_1)h(\bar{u}_1,\bar{u}_2)\period
\end{aligned}
\eeq 
The first term is the standard hexagon form factor for the magnons with rapidities $u_1$ and $u_2$. On the other hand, the second term represents a process in which the magnon $u_2$ travels to the right boundary, gets reflected and then comes back to its original position. The phase factor acquired in this process is precisely $e^{2i p_2\ell_{31}}R_{R}(u_2)$, and, since the magnon gets reflected, the relevant hexagon form factor is $h(u_1,\bar{u}_2)$. The third and the four terms can be understood in a similar manner. The only difference is that, when we move the magnon $u_1$ to the right boundary and brings it to the original position, we have to reorder the two magnons and this leads to the S-matrix factor, $S(u_1,u_2)S(u_2,\bar{u}_1)$. See also figure \ref{fig:partition} for further explanation.

\paragraph{Finite-size corrections}
As in the original hexagon proposal \cite{Basso:2015zoa}, we expect that the asymptotic answer gets corrected by the contributions from the mirror particles. In fact, our conjecture in the previous section for the BPS three-point functions consist only such mirror-particle contributions since the asymptotic part is trivial.

\begin{figure}[t]
\centering
\includegraphics[clip,height=5cm]{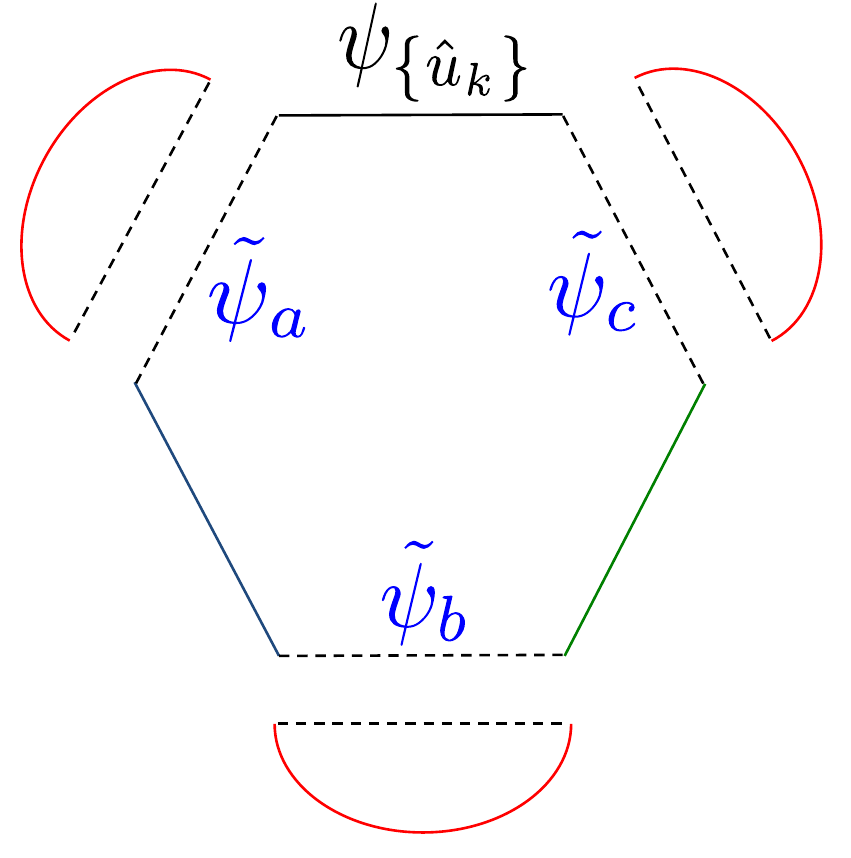}
\caption{Finite size correction to the non-BPS three-point function. The finite size correction comes from having nontrivial states on the edges glued to the boundary states. To compute it we need to consider the hexagon form factor depicted above.}
\label{fig:finite}
\end{figure}

In the case of non-BPS three-point functions, mirror particles interact with the physical magnons through the hexagon form factor. Written more explicitly, we expect that the sum over partitions $\mathcal{A}$ gets modified to (see also figure \ref{fig:finite})
\beq
\begin{aligned}
\mathcal{A}_{\text{finite-size}}=&\sum_{\alpha_{+}\cup \alpha_{-}=\{1,\dots, M\}}\left[\prod_{j\in \alpha_{-}}\left(-e^{2i p_{j}\ell_{31}}R_{R}(u_j)\right)\prod_{k>j}S(u_j,u_k)S(u_k,\bar{u}_j)\right]\\
&\times \sum_{a,b,c}e^{-(\tilde{E}_{\textcolor[rgb]{0,0,1}{\tilde{\psi}_{a}}}\ell_{12}+\tilde{E}_{\textcolor[rgb]{0,0,1}{\tilde{\psi}_{b}}}\ell_{23}+\tilde{E}_{\textcolor[rgb]{0,0,1}{\tilde{\psi}_{c}}}\ell_{31})}b_{\textcolor[rgb]{0,0,1}{\tilde{\psi}_a}}b_{\textcolor[rgb]{0,0,1}{\tilde{\psi}_b}}b_{\textcolor[rgb]{0,0,1}{\tilde{\psi}_c}}\mathcal{H}_{\textcolor[rgb]{0,0,1}{\tilde{\psi}_{a}},\textcolor[rgb]{0,0,1}{\tilde{\psi}_{b}},\textcolor[rgb]{0,0,1}{\tilde{\psi}_{c}}|\psi_{\{\hat{u}_k\}}\comma\emptyset\comma\emptyset} \comma
\end{aligned}
\eeq
with $\mathcal{H}_{\psi_1,\psi_2,\psi_3|\psi_4,\psi_5,\psi_6}$ being the hexagon form factor with $\psi_{1,2,3}$ on the mirror edges and $\psi_{4,5,6}$ on the physical edges. Here $\textcolor[rgb]{0,0,1}{\tilde{\psi}}$'s are the states obtained by the expansion of the boundary state \eqref{eq:expansionofBstate}, and $\psi_{\{\hat{u}_k\}}$ is the multi-magnon state with rapidities $\{\hat{u}_k\}$ which describes the non-BPS operator.

Of course, what we have described above is merely a conjecture since we haven't provided any  evidence apart from the tree-level analysis performed in \cite{Kim:2017phs}.
\section{Conclusion\label{sec:conclusion}}
In this paper, we computed the correlation functions on the $1/2$-BPS Wilson loop in $\mathcal{N}=4$ SYM at one loop. The results are in perfect agreement with the localization computation performed in \cite{Giombi:2018qox}. We then examined the results from the point of view of the hexagonalization approach \cite{Basso:2015zoa,Fleury:2016ykk}. 

The relation between our conjecture and the hexagon formalism for the single-trace correlators \cite{Basso:2015zoa} bears some resemblance to the KLT relation between open and closed string amplitudes in flat space \cite{Kawai:1985xq}. For instance, the number of hexagons that we need for the correlators on the Wilson loop is precisely half the number that we need for the single-trace correlators, being in line with the slogan {\it (closed string) = (open string)$^2$}. Of course the actual relation between the two is more complicated than the KLT relation since the boundary states would affect the computation nontrivially. Nevertheless it might be interesting to ask if the analogy can be made more precise.

Let us end this paper by mentioning several future directions. The most pressing question is to test our conjecture by performing the computation in the hexagonalization approach. We tried to perform such computation but did not succeed in doing so owing to the complexity of the multiparticle integrand. However, this is just technical difficulty and we believe that it can be overcome with some more efforts.

From the four-point function we computed in this paper, one can extract the structure constants of non-BPS operators on the Wilson loop by performing the conformal block expansion. A similar idea was employed in \cite{Basso:2017khq} to test the hexagon formalism for the single-trace structure constants. It would be interesting to perform the same analysis in our set-up and check the proposal in \cite{Kim:2017phs} at one loop and extend it to the higher-rank sectors.

Another interesting direction is to study the strong-coupling limit and compare it with the results in \cite{Giombi:2017cqn}. Also interesting would be to consider long operators at strong coupling. Such operators are dual to the classical open strings and the correlators are described by some nontrivial classical string configuration. Determining the shape of such a configuration is generally a hard problem, but one may be able to compute its area directly without needing to know its shape by applying the methods developed in \cite{Caetano:2012ac,Janik:2011bd,Kazama:2016cfl,Kazama:2013qsa}. It would also be interesting to look for the analogue of the ``simple'' four-point functions analyzed in \cite{Coronado:2018ypq,Coronado:2018cxj}.

We would also like to emphasize that the correlators on the Wilson loop are ideal quantities for understanding the connection between the OPE and integrability: Unlike standard single-trace operators whose OPE necessarily involves the multi-trace operators even in the planar limit, the OPE of operators on the Wilson loop is closed at large $N$. This would make it simpler to compare the OPE with the integrability-based approaches, and understand the relation between the two  \cite{Basso:2017khq}.

It goes without saying that one can try to generalize our analysis to other set-ups which are also described by the open string worldsheet. It would be particularly interesting to study the correlators of mesonic operators in the flavored/orientifolded $\mathcal{N}=4$ SYM \cite{Chen:2004mu,Erler:2005nr,Correa:2008av}\fn{Other possible set-ups are the correlators on the domain-wall defect \cite{DeWolfe:2004zt,Okamura:2006zr} and the correlators of determinant-like operators which are described by open strings on the $Y=0$ branes \cite{Berenstein:2005vf,Hofman:2007xp,Bajnok:2012xc}.}.  

It would also be interesting to generalize our perturbative analysis to the non-planar level and check the predictions for the non-planar correlators from the localization in \cite{Giombi:2018hsx}.

More ambitiously, it would be exciting if one could express the correlators in the Chern-Simons vector models in three dimensions \cite{Giombi:2009wh,Giombi:2011kc,Aharony:2012nh,Turiaci:2018nua} in terms of hexagons. Being vector-like models, they likely admit a description in terms of some (limit of) open string theory in AdS\fn{See \cite{Chang:2012kt} for explicit proposals on the open-string dual of some of the vector-like models.}, and one might hope that their correlators can be decomposed in a way similar to the ones proposed in this paper. If true, that would provide a clue to the worldsheet description of the Vasiliev theory \cite{Giombi:2009wh,Vasiliev:1995dn,Klebanov:2002ja} to which these theories are believed to be dual to. This was actually the original motivation for this work and we hope to come back to it in the not-so-distant future.
\section*{Acknowledgement}
We thank Thiago Fleury for sharing with us the Mathematica notebooks on the computation of the matrix part of the hexagon form factor.
We are grateful to Perimeter Institute for Theoretical Physics where this work was initiated.
The research of NK is supported in part
by JSPS Research Fellowship for Young Scientists, from the Japan Ministry of Education,
Culture, Sports, Science and Technology while the research of SK is
supported by DOE grant number DE-SC0009988.
\appendix
\section{Action and Propagators\label{ap:action}}
Here we summarize our convention of the action and the propagators. The convention follows the one in \cite{Kim:2017sju}, which is equivalent to the one in \cite{Erickson:2000af} up to minor modifications.

In our convention, the action of $\mathcal{N}=4$ SYM in the Euclidean signature reads
\begin{align}
&S=\frac{1}{g_{\rm YM}^2}\int d^{4} x\,\,\mathcal{L}\comma\\
&\mathcal{L}={\rm Tr}\left[-\frac{[D_{\mu},D_{\nu}]^2}{2}+(D_{\mu}\phi_i)^2+\frac{[\phi_i,\phi_j]^2}{2}+i\bar{\psi}\Gamma^{\mu}D_{\mu}\psi+\bar{\psi}\Gamma^{i}[\phi_i,\psi]+\del^{\mu}\bar{c}D_{\mu}c+(\del_{\mu}A^{\mu})^2\right]\comma\nonumber
\end{align}
where $D_{\mu}\equiv \del_{\mu}-i[A_{\mu}, \bullet]$ and $c$ and $\bar{c}$ are the BRST ghosts. The Gamma matrices $\Gamma^{A}=(\Gamma^{\mu},\Gamma^{i})$ are the ten-dimensional Dirac matrices which satisfy the orthogonality relation
\beq
{\rm tr}\left(\Gamma^{A}\Gamma^{B}\right)=16\delta^{AB}\period
\eeq 

The propagators of individual fields can be read off from this action. For instance, the results for the gauge field and the scalar read
\beq
\begin{aligned}
\langle (A_{\mu})^{a}{}_{b}(x) (A_{\nu})^{c}{}_{d}(y)\rangle&=\frac{g_{\rm YM}^2\delta^{a}{}_{c}\delta^{c}{}_{b}}{8\pi^2}\frac{\delta_{\mu\nu}}{|x-y|^2}\comma\\
\langle (\phi_i)^{a}{}_{b}(x)(\phi_j)^{c}{}_{d}(y)\rangle&=\frac{g_{\rm YM}^2\delta^{a}{}_{c}\delta^{c}{}_{b}}{8\pi^2}\frac{\delta_{ij}}{|x-y|^2}\comma
\end{aligned}
\eeq
where $a$-$d$ are the color indices.
\bibliographystyle{utphys}
\bibliography{4ptwlref}

\providecommand{\href}[2]{#2}\begingroup\raggedright\begin{thebibliography}{10}

\bibitem{Polyakov:1980ca}
A.~M. Polyakov, ``{Gauge Fields as Rings of Glue},''
\href{http://dx.doi.org/10.1016/0550-3213(80)90507-6}{{\em Nucl. Phys.}
  {\bfseries B164} (1980) 171--188}.

\bibitem{Makeenko:1979pb}
{\relax Yu}.~M. Makeenko and A.~A. Migdal, ``{Exact Equation for the Loop
  Average in Multicolor QCD},''
  \href{http://dx.doi.org/10.1016/0370-2693(79)90131-X}{{\em Phys. Lett.}
  {\bfseries 88B} (1979) 135}.
[Erratum: Phys. Lett.89B,437(1980)].

\bibitem{Makeenko:1980vm}
{\relax Yu}.~Makeenko and A.~A. Migdal, ``{Quantum Chromodynamics as Dynamics
  of Loops},'' \href{http://dx.doi.org/10.1016/0550-3213(81)90258-3}{{\em Nucl.
  Phys.} {\bfseries B188} (1981) 269}.
[Yad. Fiz.32,838(1980)].

\bibitem{Maldacena:1997re}
J.~M. Maldacena, ``{The Large N limit of superconformal field theories and
  supergravity},'' \href{http://dx.doi.org/10.1023/A:1026654312961,
  10.4310/ATMP.1998.v2.n2.a1}{{\em Int. J. Theor. Phys.} {\bfseries 38} (1999)
  1113--1133}, \href{http://arxiv.org/abs/hep-th/9711200}{{\ttfamily
  arXiv:hep-th/9711200 [hep-th]}}.
[Adv. Theor. Math. Phys.2,231(1998)].

\bibitem{tHooft:1973alw}
G.~'t~Hooft, ``{A Planar Diagram Theory for Strong Interactions},''
  \href{http://dx.doi.org/10.1016/0550-3213(74)90154-0}{{\em Nucl. Phys.}
  {\bfseries B72} (1974) 461}.
[,337(1973)].

\bibitem{Minahan:2002ve}
J.~A. Minahan and K.~Zarembo, ``{The Bethe ansatz for N=4 superYang-Mills},''
  \href{http://dx.doi.org/10.1088/1126-6708/2003/03/013}{{\em JHEP} {\bfseries
  03} (2003) 013},
\href{http://arxiv.org/abs/hep-th/0212208}{{\ttfamily arXiv:hep-th/0212208
  [hep-th]}}.

\bibitem{Beisert:2010jr}
N.~Beisert {\em et~al.}, ``{Review of AdS/CFT Integrability: An Overview},''
  \href{http://dx.doi.org/10.1007/s11005-011-0529-2}{{\em Lett. Math. Phys.}
  {\bfseries 99} (2012) 3--32},
\href{http://arxiv.org/abs/1012.3982}{{\ttfamily arXiv:1012.3982 [hep-th]}}.

\bibitem{Gromov:2009tv}
N.~Gromov, V.~Kazakov, and P.~Vieira, ``{Exact Spectrum of Anomalous Dimensions
  of Planar N=4 Supersymmetric Yang-Mills Theory},''
  \href{http://dx.doi.org/10.1103/PhysRevLett.103.131601}{{\em Phys. Rev.
  Lett.} {\bfseries 103} (2009) 131601},
\href{http://arxiv.org/abs/0901.3753}{{\ttfamily arXiv:0901.3753 [hep-th]}}.

\bibitem{Bombardelli:2009ns}
D.~Bombardelli, D.~Fioravanti, and R.~Tateo, ``{Thermodynamic Bethe Ansatz for
  planar AdS/CFT: A Proposal},''
  \href{http://dx.doi.org/10.1088/1751-8113/42/37/375401}{{\em J. Phys.}
  {\bfseries A42} (2009) 375401},
\href{http://arxiv.org/abs/0902.3930}{{\ttfamily arXiv:0902.3930 [hep-th]}}.

\bibitem{Arutyunov:2009ur}
G.~Arutyunov and S.~Frolov, ``{Thermodynamic Bethe Ansatz for the AdS(5) x S(5)
  Mirror Model},'' \href{http://dx.doi.org/10.1088/1126-6708/2009/05/068}{{\em
  JHEP} {\bfseries 05} (2009) 068},
\href{http://arxiv.org/abs/0903.0141}{{\ttfamily arXiv:0903.0141 [hep-th]}}.

\bibitem{Gromov:2013pga}
N.~Gromov, V.~Kazakov, S.~Leurent, and D.~Volin, ``{Quantum Spectral Curve for
  Planar $\mathcal{N} = 4$ Super-Yang-Mills Theory},''
  \href{http://dx.doi.org/10.1103/PhysRevLett.112.011602}{{\em Phys. Rev.
  Lett.} {\bfseries 112} no.~1, (2014) 011602},
\href{http://arxiv.org/abs/1305.1939}{{\ttfamily arXiv:1305.1939 [hep-th]}}.

\bibitem{Basso:2013vsa}
B.~Basso, A.~Sever, and P.~Vieira, ``{Spacetime and Flux Tube S-Matrices at
  Finite Coupling for N=4 Supersymmetric Yang-Mills Theory},''
  \href{http://dx.doi.org/10.1103/PhysRevLett.111.091602}{{\em Phys. Rev.
  Lett.} {\bfseries 111} no.~9, (2013) 091602},
\href{http://arxiv.org/abs/1303.1396}{{\ttfamily arXiv:1303.1396 [hep-th]}}.

\bibitem{Basso:2015zoa}
B.~Basso, S.~Komatsu, and P.~Vieira, ``{Structure Constants and Integrable
  Bootstrap in Planar N=4 SYM Theory},''
\href{http://arxiv.org/abs/1505.06745}{{\ttfamily arXiv:1505.06745 [hep-th]}}.

\bibitem{Caetano:2011eb}
J.~Caetano and J.~Escobedo, ``{On four-point functions and integrability in N=4
  SYM: from weak to strong coupling},''
  \href{http://dx.doi.org/10.1007/JHEP09(2011)080}{{\em JHEP} {\bfseries 09}
  (2011) 080},
\href{http://arxiv.org/abs/1107.5580}{{\ttfamily arXiv:1107.5580 [hep-th]}}.

\bibitem{Eden:2016xvg}
B.~Eden and A.~Sfondrini, ``{Tessellating cushions: four-point functions in
  $\mathcal{N} $ = 4 SYM},''
  \href{http://dx.doi.org/10.1007/JHEP10(2017)098}{{\em JHEP} {\bfseries 10}
  (2017) 098},
\href{http://arxiv.org/abs/1611.05436}{{\ttfamily arXiv:1611.05436 [hep-th]}}.

\bibitem{Caetano:2012ac}
J.~Caetano and J.~Toledo, ``{$\chi$-Systems for Correlation Functions},''
\href{http://arxiv.org/abs/1208.4548}{{\ttfamily arXiv:1208.4548 [hep-th]}}.

\bibitem{Fleury:2016ykk}
T.~Fleury and S.~Komatsu, ``{Hexagonalization of Correlation Functions},''
  \href{http://dx.doi.org/10.1007/JHEP01(2017)130}{{\em JHEP} {\bfseries 01}
  (2017) 130},
\href{http://arxiv.org/abs/1611.05577}{{\ttfamily arXiv:1611.05577 [hep-th]}}.

\bibitem{Fleury:2017eph}
T.~Fleury and S.~Komatsu, ``{Hexagonalization of Correlation Functions II:
  Two-Particle Contributions},''
  \href{http://dx.doi.org/10.1007/JHEP02(2018)177}{{\em JHEP} {\bfseries 02}
  (2018) 177},
\href{http://arxiv.org/abs/1711.05327}{{\ttfamily arXiv:1711.05327 [hep-th]}}.

\bibitem{Ben-Israel:2018ckc}
R.~Ben-Israel, A.~G. Tumanov, and A.~Sever, ``{Scattering amplitudes — Wilson
  loops duality for the first non-planar correction},''
  \href{http://dx.doi.org/10.1007/JHEP08(2018)122}{{\em JHEP} {\bfseries 08}
  (2018) 122},
\href{http://arxiv.org/abs/1802.09395}{{\ttfamily arXiv:1802.09395 [hep-th]}}.

\bibitem{Bargheer:2017nne}
T.~Bargheer, J.~Caetano, T.~Fleury, S.~Komatsu, and P.~Vieira, ``{Handling
  Handles I: Nonplanar Integrability},''
\href{http://arxiv.org/abs/1711.05326}{{\ttfamily arXiv:1711.05326 [hep-th]}}.

\bibitem{Eden:2017ozn}
B.~Eden, Y.~Jiang, D.~le~Plat, and A.~Sfondrini, ``{Colour-dressed hexagon
  tessellations for correlation functions and non-planar corrections},''
  \href{http://dx.doi.org/10.1007/JHEP02(2018)170}{{\em JHEP} {\bfseries 02}
  (2018) 170},
\href{http://arxiv.org/abs/1710.10212}{{\ttfamily arXiv:1710.10212 [hep-th]}}.

\bibitem{Bargheer:2018jvq}
T.~Bargheer, J.~Caetano, T.~Fleury, S.~Komatsu, and P.~Vieira, ``{Handling
  Handles II: Stratification and Data Analysis},''
\href{http://arxiv.org/abs/1809.09145}{{\ttfamily arXiv:1809.09145 [hep-th]}}.

\bibitem{Erickson:2000af}
J.~K. Erickson, G.~W. Semenoff, and K.~Zarembo, ``{Wilson loops in N=4
  supersymmetric Yang-Mills theory},''
  \href{http://dx.doi.org/10.1016/S0550-3213(00)00300-X}{{\em Nucl. Phys.}
  {\bfseries B582} (2000) 155--175},
\href{http://arxiv.org/abs/hep-th/0003055}{{\ttfamily arXiv:hep-th/0003055
  [hep-th]}}.

\bibitem{Drukker:2000rr}
N.~Drukker and D.~J. Gross, ``{An Exact prediction of N=4 SUSYM theory for
  string theory},'' \href{http://dx.doi.org/10.1063/1.1372177}{{\em J. Math.
  Phys.} {\bfseries 42} (2001) 2896--2914},
\href{http://arxiv.org/abs/hep-th/0010274}{{\ttfamily arXiv:hep-th/0010274
  [hep-th]}}.

\bibitem{Pestun:2007rz}
V.~Pestun, ``{Localization of gauge theory on a four-sphere and supersymmetric
  Wilson loops},'' \href{http://dx.doi.org/10.1007/s00220-012-1485-0}{{\em
  Commun. Math. Phys.} {\bfseries 313} (2012) 71--129},
\href{http://arxiv.org/abs/0712.2824}{{\ttfamily arXiv:0712.2824 [hep-th]}}.

\bibitem{Billo:2016cpy}
M.~Billò, V.~Gonçalves, E.~Lauria, and M.~Meineri, ``{Defects in conformal
  field theory},'' \href{http://dx.doi.org/10.1007/JHEP04(2016)091}{{\em JHEP}
  {\bfseries 04} (2016) 091},
\href{http://arxiv.org/abs/1601.02883}{{\ttfamily arXiv:1601.02883 [hep-th]}}.

\bibitem{Liendo:2016ymz}
P.~Liendo and C.~Meneghelli, ``{Bootstrap equations for $ \mathcal{N} $ = 4 SYM
  with defects},'' \href{http://dx.doi.org/10.1007/JHEP01(2017)122}{{\em JHEP}
  {\bfseries 01} (2017) 122},
\href{http://arxiv.org/abs/1608.05126}{{\ttfamily arXiv:1608.05126 [hep-th]}}.

\bibitem{Liendo:2018ukf}
P.~Liendo, C.~Meneghelli, and V.~Mitev, ``{Bootstrapping the half-BPS line
  defect},'' \href{http://dx.doi.org/10.1007/JHEP10(2018)077}{{\em JHEP}
  {\bfseries 10} (2018) 077},
\href{http://arxiv.org/abs/1806.01862}{{\ttfamily arXiv:1806.01862 [hep-th]}}.

\bibitem{Kim:2017sju}
M.~Kim, N.~Kiryu, S.~Komatsu, and T.~Nishimura, ``{Structure Constants of
  Defect Changing Operators on the 1/2 BPS Wilson Loop},''
  \href{http://dx.doi.org/10.1007/JHEP12(2017)055}{{\em JHEP} {\bfseries 12}
  (2017) 055},
\href{http://arxiv.org/abs/1710.07325}{{\ttfamily arXiv:1710.07325 [hep-th]}}.

\bibitem{Cavaglia:2018lxi}
A.~Cavaglia, N.~Gromov, and F.~Levkovich-Maslyuk, ``{Quantum Spectral Curve and
  Structure Constants in N=4 SYM: Cusps in the Ladder Limit},''
\href{http://arxiv.org/abs/1802.04237}{{\ttfamily arXiv:1802.04237 [hep-th]}}.

\bibitem{Giombi:2018qox}
S.~Giombi and S.~Komatsu, ``{Exact Correlators on the Wilson Loop in
  $\mathcal{N}=4$ SYM: Localization, Defect CFT, and Integrability},''
  \href{http://dx.doi.org/10.1007/JHEP05(2018)109}{{\em JHEP} {\bfseries 05}
  (2018) 109},
\href{http://arxiv.org/abs/1802.05201}{{\ttfamily arXiv:1802.05201 [hep-th]}}.

\bibitem{Giombi:2018hsx}
S.~Giombi and S.~Komatsu, ``{More Exact Results in the Wilson Loop Defect CFT:
  Bulk-Defect OPE, Nonplanar Corrections and Quantum Spectral Curve},''
\href{http://arxiv.org/abs/1811.02369}{{\ttfamily arXiv:1811.02369 [hep-th]}}.

\bibitem{Beccaria:2017rbe}
M.~Beccaria, S.~Giombi, and A.~Tseytlin, ``{Non-supersymmetric Wilson loop in $
  \mathcal{N} $ = 4 SYM and defect 1d CFT},''
  \href{http://dx.doi.org/10.1007/JHEP03(2018)131}{{\em JHEP} {\bfseries 03}
  (2018) 131},
\href{http://arxiv.org/abs/1712.06874}{{\ttfamily arXiv:1712.06874 [hep-th]}}.

\bibitem{Beccaria:2018ocq}
M.~Beccaria and A.~A. Tseytlin, ``{On non-supersymmetric generalizations of the
  Wilson-Maldacena loops in $N=4$ SYM},''
  \href{http://dx.doi.org/10.1016/j.nuclphysb.2018.07.019}{{\em Nucl. Phys.}
  {\bfseries B934} (2018) 466--497},
\href{http://arxiv.org/abs/1804.02179}{{\ttfamily arXiv:1804.02179 [hep-th]}}.

\bibitem{Correa:2018fgz}
D.~Correa, M.~Leoni, and S.~Luque, ``{Spin chain integrability in
  non-supersymmetric Wilson loops},''
\href{http://arxiv.org/abs/1810.04643}{{\ttfamily arXiv:1810.04643 [hep-th]}}.

\bibitem{Giombi:2017cqn}
S.~Giombi, R.~Roiban, and A.~A. Tseytlin, ``{Half-BPS Wilson loop and
  AdS$_2$/CFT$_1$},''
  \href{http://dx.doi.org/10.1016/j.nuclphysb.2017.07.004}{{\em Nucl. Phys.}
  {\bfseries B922} (2017) 499--527},
\href{http://arxiv.org/abs/1706.00756}{{\ttfamily arXiv:1706.00756 [hep-th]}}.

\bibitem{Cooke:2017qgm}
M.~Cooke, A.~Dekel, and N.~Drukker, ``{The Wilson loop CFT: Insertion
  dimensions and structure constants from wavy lines},''
  \href{http://dx.doi.org/10.1088/1751-8121/aa7db4}{{\em J. Phys.} {\bfseries
  A50} no.~33, (2017) 335401},
\href{http://arxiv.org/abs/1703.03812}{{\ttfamily arXiv:1703.03812 [hep-th]}}.

\bibitem{Cooke:2018obg}
M.~Cooke, A.~Dekel, N.~Drukker, D.~Trancanelli, and E.~Vescovi, ``{Deformations
  of the circular Wilson loop and spectral (in)dependence},''
\href{http://arxiv.org/abs/1811.09638}{{\ttfamily arXiv:1811.09638 [hep-th]}}.

\bibitem{Drukker:2006xg}
N.~Drukker and S.~Kawamoto, ``{Small deformations of supersymmetric Wilson
  loops and open spin-chains},''
  \href{http://dx.doi.org/10.1088/1126-6708/2006/07/024}{{\em JHEP} {\bfseries
  07} (2006) 024},
\href{http://arxiv.org/abs/hep-th/0604124}{{\ttfamily arXiv:hep-th/0604124
  [hep-th]}}.

\bibitem{Drukker:2012de}
N.~Drukker, ``{Integrable Wilson loops},''
  \href{http://dx.doi.org/10.1007/JHEP10(2013)135}{{\em JHEP} {\bfseries 10}
  (2013) 135},
\href{http://arxiv.org/abs/1203.1617}{{\ttfamily arXiv:1203.1617 [hep-th]}}.

\bibitem{Correa:2012hh}
D.~Correa, J.~Maldacena, and A.~Sever, ``{The quark anti-quark potential and
  the cusp anomalous dimension from a TBA equation},''
  \href{http://dx.doi.org/10.1007/JHEP08(2012)134}{{\em JHEP} {\bfseries 08}
  (2012) 134},
\href{http://arxiv.org/abs/1203.1913}{{\ttfamily arXiv:1203.1913 [hep-th]}}.

\bibitem{Gromov:2015dfa}
N.~Gromov and F.~Levkovich-Maslyuk, ``{Quantum Spectral Curve for a cusped
  Wilson line in $ \mathcal{N}=4 $ SYM},''
  \href{http://dx.doi.org/10.1007/JHEP04(2016)134}{{\em JHEP} {\bfseries 04}
  (2016) 134},
\href{http://arxiv.org/abs/1510.02098}{{\ttfamily arXiv:1510.02098 [hep-th]}}.

\bibitem{Kim:2017phs}
M.~Kim and N.~Kiryu, ``{Structure constants of operators on the Wilson loop
  from integrability},'' \href{http://dx.doi.org/10.1007/JHEP11(2017)116}{{\em
  JHEP} {\bfseries 11} (2017) 116},
\href{http://arxiv.org/abs/1706.02989}{{\ttfamily arXiv:1706.02989 [hep-th]}}.

\bibitem{Drukker:2008pi}
N.~Drukker and J.~Plefka, ``{The Structure of n-point functions of chiral
  primary operators in N=4 super Yang-Mills at one-loop},''
  \href{http://dx.doi.org/10.1088/1126-6708/2009/04/001}{{\em JHEP} {\bfseries
  04} (2009) 001},
\href{http://arxiv.org/abs/0812.3341}{{\ttfamily arXiv:0812.3341 [hep-th]}}.

\bibitem{Hellerman:2015nra}
S.~Hellerman, D.~Orlando, S.~Reffert, and M.~Watanabe, ``{On the CFT Operator
  Spectrum at Large Global Charge},''
  \href{http://dx.doi.org/10.1007/JHEP12(2015)071}{{\em JHEP} {\bfseries 12}
  (2015) 071},
\href{http://arxiv.org/abs/1505.01537}{{\ttfamily arXiv:1505.01537 [hep-th]}}.

\bibitem{Kawai:1985xq}
H.~Kawai, D.~C. Lewellen, and S.~H.~H. Tye, ``{A Relation Between Tree
  Amplitudes of Closed and Open Strings},''
\href{http://dx.doi.org/10.1016/0550-3213(86)90362-7}{{\em Nucl. Phys.}
  {\bfseries B269} (1986) 1--23}.

\bibitem{Basso:2017khq}
B.~Basso, F.~Coronado, S.~Komatsu, H.~T. Lam, P.~Vieira, and D.-l. Zhong,
  ``{Asymptotic Four Point Functions},''
\href{http://arxiv.org/abs/1701.04462}{{\ttfamily arXiv:1701.04462 [hep-th]}}.

\bibitem{Janik:2011bd}
R.~A. Janik and A.~Wereszczynski, ``{Correlation functions of three heavy
  operators: The AdS contribution},''
  \href{http://dx.doi.org/10.1007/JHEP12(2011)095}{{\em JHEP} {\bfseries 12}
  (2011) 095},
\href{http://arxiv.org/abs/1109.6262}{{\ttfamily arXiv:1109.6262 [hep-th]}}.

\bibitem{Kazama:2016cfl}
Y.~Kazama, S.~Komatsu, and T.~Nishimura, ``{Classical Integrability for
  Three-point Functions: Cognate Structure at Weak and Strong Couplings},''
  \href{http://dx.doi.org/10.1007/JHEP10(2016)042,
  10.1007/JHEP02(2018)047}{{\em JHEP} {\bfseries 10} (2016) 042},
  \href{http://arxiv.org/abs/1603.03164}{{\ttfamily arXiv:1603.03164
  [hep-th]}}.
[Erratum: JHEP02,047(2018)].

\bibitem{Kazama:2013qsa}
Y.~Kazama and S.~Komatsu, ``{Three-point functions in the SU(2) sector at
  strong coupling},'' \href{http://dx.doi.org/10.1007/JHEP03(2014)052}{{\em
  JHEP} {\bfseries 03} (2014) 052},
\href{http://arxiv.org/abs/1312.3727}{{\ttfamily arXiv:1312.3727 [hep-th]}}.

\bibitem{Coronado:2018ypq}
F.~Coronado, ``{Perturbative Four-Point Functions In Planar N=4 SYM From
  Hexagonalization},''
\href{http://arxiv.org/abs/1811.00467}{{\ttfamily arXiv:1811.00467 [hep-th]}}.

\bibitem{Coronado:2018cxj}
F.~Coronado, ``{Bootstrapping the simplest correlator in planar N = 4 SYM at
  all loops},''
\href{http://arxiv.org/abs/1811.03282}{{\ttfamily arXiv:1811.03282 [hep-th]}}.

\bibitem{Chen:2004mu}
B.~Chen, X.-J. Wang, and Y.-S. Wu, ``{Integrable open spin chain in
  superYang-Mills and the plane wave / SYM duality},''
  \href{http://dx.doi.org/10.1088/1126-6708/2004/02/029}{{\em JHEP} {\bfseries
  02} (2004) 029},
\href{http://arxiv.org/abs/hep-th/0401016}{{\ttfamily arXiv:hep-th/0401016
  [hep-th]}}.

\bibitem{Erler:2005nr}
T.~Erler and N.~Mann, ``{Integrable open spin chains and the doubling trick in
  N=2 SYM with fundamental matter},''
  \href{http://dx.doi.org/10.1088/1126-6708/2006/01/131}{{\em JHEP} {\bfseries
  01} (2006) 131},
\href{http://arxiv.org/abs/hep-th/0508064}{{\ttfamily arXiv:hep-th/0508064
  [hep-th]}}.

\bibitem{Correa:2008av}
D.~H. Correa and C.~A.~S. Young, ``{Reflecting magnons from D7 and D5
  branes},'' \href{http://dx.doi.org/10.1088/1751-8113/41/45/455401}{{\em J.
  Phys.} {\bfseries A41} (2008) 455401},
\href{http://arxiv.org/abs/0808.0452}{{\ttfamily arXiv:0808.0452 [hep-th]}}.

\bibitem{DeWolfe:2004zt}
O.~DeWolfe and N.~Mann, ``{Integrable open spin chains in defect conformal
  field theory},'' \href{http://dx.doi.org/10.1088/1126-6708/2004/04/035}{{\em
  JHEP} {\bfseries 04} (2004) 035},
\href{http://arxiv.org/abs/hep-th/0401041}{{\ttfamily arXiv:hep-th/0401041
  [hep-th]}}.

\bibitem{Okamura:2006zr}
K.~Okamura and K.~Yoshida, ``{Higher Loop Bethe Ansatz for Open Spin-Chains in
  AdS/CFT},'' \href{http://dx.doi.org/10.1088/1126-6708/2006/09/081}{{\em JHEP}
  {\bfseries 09} (2006) 081},
\href{http://arxiv.org/abs/hep-th/0604100}{{\ttfamily arXiv:hep-th/0604100
  [hep-th]}}.

\bibitem{Berenstein:2005vf}
D.~Berenstein and S.~E. Vazquez, ``{Integrable open spin chains from giant
  gravitons},'' \href{http://dx.doi.org/10.1088/1126-6708/2005/06/059}{{\em
  JHEP} {\bfseries 06} (2005) 059},
\href{http://arxiv.org/abs/hep-th/0501078}{{\ttfamily arXiv:hep-th/0501078
  [hep-th]}}.

\bibitem{Hofman:2007xp}
D.~M. Hofman and J.~M. Maldacena, ``{Reflecting magnons},''
  \href{http://dx.doi.org/10.1088/1126-6708/2007/11/063}{{\em JHEP} {\bfseries
  11} (2007) 063},
\href{http://arxiv.org/abs/0708.2272}{{\ttfamily arXiv:0708.2272 [hep-th]}}.

\bibitem{Bajnok:2012xc}
Z.~Bajnok, R.~I. Nepomechie, L.~Palla, and R.~Suzuki, ``{Y-system for Y=0 brane
  in planar AdS/CFT},'' \href{http://dx.doi.org/10.1007/JHEP08(2012)149}{{\em
  JHEP} {\bfseries 08} (2012) 149},
\href{http://arxiv.org/abs/1205.2060}{{\ttfamily arXiv:1205.2060 [hep-th]}}.

\bibitem{Giombi:2009wh}
S.~Giombi and X.~Yin, ``{Higher Spin Gauge Theory and Holography: The
  Three-Point Functions},''
  \href{http://dx.doi.org/10.1007/JHEP09(2010)115}{{\em JHEP} {\bfseries 09}
  (2010) 115},
\href{http://arxiv.org/abs/0912.3462}{{\ttfamily arXiv:0912.3462 [hep-th]}}.

\bibitem{Giombi:2011kc}
S.~Giombi, S.~Minwalla, S.~Prakash, S.~P. Trivedi, S.~R. Wadia, and X.~Yin,
  ``{Chern-Simons Theory with Vector Fermion Matter},''
  \href{http://dx.doi.org/10.1140/epjc/s10052-012-2112-0}{{\em Eur. Phys. J.}
  {\bfseries C72} (2012) 2112},
\href{http://arxiv.org/abs/1110.4386}{{\ttfamily arXiv:1110.4386 [hep-th]}}.

\bibitem{Aharony:2012nh}
O.~Aharony, G.~Gur-Ari, and R.~Yacoby, ``{Correlation Functions of Large N
  Chern-Simons-Matter Theories and Bosonization in Three Dimensions},''
  \href{http://dx.doi.org/10.1007/JHEP12(2012)028}{{\em JHEP} {\bfseries 12}
  (2012) 028},
\href{http://arxiv.org/abs/1207.4593}{{\ttfamily arXiv:1207.4593 [hep-th]}}.

\bibitem{Turiaci:2018nua}
G.~J. Turiaci and A.~Zhiboedov, ``{Veneziano Amplitude of Vasiliev Theory},''
  \href{http://dx.doi.org/10.1007/JHEP10(2018)034}{{\em JHEP} {\bfseries 10}
  (2018) 034},
\href{http://arxiv.org/abs/1802.04390}{{\ttfamily arXiv:1802.04390 [hep-th]}}.

\bibitem{Chang:2012kt}
C.-M. Chang, S.~Minwalla, T.~Sharma, and X.~Yin, ``{ABJ Triality: from Higher
  Spin Fields to Strings},''
  \href{http://dx.doi.org/10.1088/1751-8113/46/21/214009}{{\em J. Phys.}
  {\bfseries A46} (2013) 214009},
\href{http://arxiv.org/abs/1207.4485}{{\ttfamily arXiv:1207.4485 [hep-th]}}.

\bibitem{Vasiliev:1995dn}
M.~A. Vasiliev, ``{Higher spin gauge theories in four-dimensions,
  three-dimensions, and two-dimensions},''
  \href{http://dx.doi.org/10.1142/S0218271896000473}{{\em Int. J. Mod. Phys.}
  {\bfseries D5} (1996) 763--797},
\href{http://arxiv.org/abs/hep-th/9611024}{{\ttfamily arXiv:hep-th/9611024
  [hep-th]}}.

\bibitem{Klebanov:2002ja}
I.~R. Klebanov and A.~M. Polyakov, ``{AdS dual of the critical O(N) vector
  model},'' \href{http://dx.doi.org/10.1016/S0370-2693(02)02980-5}{{\em Phys.
  Lett.} {\bfseries B550} (2002) 213--219},
\href{http://arxiv.org/abs/hep-th/0210114}{{\ttfamily arXiv:hep-th/0210114
  [hep-th]}}.

\end{thebibliography}\endgroup
\end{document}